\documentclass[epj]{svjour}
\usepackage{comment}
\usepackage{graphicx}
\usepackage{dcolumn}
\usepackage{bm}
\usepackage{subcaption}
\usepackage{authblk}
\usepackage{tablefootnote}

\usepackage{hyperref}
\hypersetup{
    colorlinks,
    citecolor=red,
    filecolor=green,
    linkcolor=blue,
    urlcolor=blue
}

\begin{document}

\title{Triplet Lifetime in Gaseous Argon}
 
\author{Michael Akashi-Ronquest\inst{5,12} \and Amanda Bacon\inst{1} \and Christopher Benson\inst{4,10} \and Kolahal Bhattacharya\inst{7} \and Thomas Caldwell\inst{12,13} \and Joseph A. Formaggio \inst{6} \and Dan Gastler\inst{2} \and Brianna Grado-White\inst{4,10} \and Jeff Griego\inst{5} \and Michael Gold\inst{11} \and Andrew Hime\inst{7} \and Christopher M. Jackson\inst{4,7,10} \and Stephen Jaditz\inst{6} \and Chris Kachulis\inst{2} \and Edward Kearns\inst{2}  \and  Joshua R.  Klein\inst{13} \and Antonio Ledesma\inst{7} \and Steve Linden\inst{2} \and Frank Lopez\inst{5} \and  Sean MacMullin\inst{12} \and Andrew Mastbaum\inst{13} \and Jocelyn Monroe\inst{6,8} \and James Nikkel\inst{15} \and  John Oertel \inst{5}  \and Gabriel D. Orebi Gann\inst{4,10} \and Gabriel S. Ortega\inst{7} \and Kimberley Palladino\inst{6,9,14} \and Keith Rielage\inst{5} \and Stanley R. Seibert\inst{5,13} \and Jui-Jen Wang\inst{3,11} 
\thanks{\emph{Corresponding author : } wangbtc@brandeis.edu}%
} 

\institute{Bennington College, Bennington, VT 05201, USA \and Department of Physics, Boston University, Boston, MA 02215, USA \and Department of Physics, Brandeis University, Waltham, MA 02453, USA\and Lawrence Berkeley National Laboratory, Berkeley, CA 94720, USA \and Los Alamos National Laboratory, Los Alamos, NM 87545, USA \and Laboratory for Nuclear Science, Massachusetts Institute of Technology, Cambridge, MA 02139, USA \and Pacific Northwest National Laboratory, Richland, WA 99352, USA \and Department of Physics, Royal Holloway, University of London, Egham TW20 0EX, UK\and SNOLAB Institute, Lively, ON P3Y 1N2, Canada \and Department of Physics, University of California, Berkeley, Berkeley, CA 94720, USA \and Department of Physics and Astronomy, University of New Mexico, Albuquerque, NM 87131, USA \and Department of Physics and Astronomy, University of North Carolina, Chapel Hill, NC 27599, USA \and Department of Physics and Astronomy, University of Pennsylvania, Philadelphia, PA 19104, USA \and Department of Physics, University of Wisconsin-Madison, Madison, WI 53706, USA \and Department of Physics, Yale University, New Haven, CT 06511, USA}


\date{\today}

\abstract{
  MiniCLEAN is a single-phase liquid argon dark matter experiment.
 During the initial cooling phase, impurities within the cold gas ($<$140 K) were monitored by measuring the scintillation light triplet lifetime,
 and ultimately a triplet lifetime of 3.480 $\pm$ 0.001 (stat.) $\pm$ 0.064 (sys.) $\mu$s was obtained, indicating ultra-pure argon.  
 This is the longest argon triplet time constant ever reported. The effect of quenching 
of separate components of the scintillation light is also investigated.
\PACS{
      {PACS-key}{discribing text of that key}   \and
      {PACS-key}{discribing text of that key}
     } 
} 
\maketitle

\section{Introduction}
\label{S:1}

Argon, Neon and Xenon noble liquids are of extensive interest to both direct dark matter searches and neutrino physics experiments. The high scintillation light yield of argon in particular gives relatively good energy resolution and its low ionization potential makes it suitable for charge detection. In addition, argon is well-suited for building large volume detectors due to its relatively low cost. This paper presents the measurement of gaseous argon properties while the MiniCLEAN detector was in its cooling phase.  MiniCLEAN is a prototype for the full scale CLEAN (Cryogenic Low-Energy Astrophysics with Noble liquids) program which is a single-phase liquid argon dark matter experiment.

\par
In gaseous argon, the spectrum of scintillation light is produced in three continuous bands. The first continuum ranges from 104~nm to 110~nm, the second continuum peaks at 128~nm and the third continuum ranges from 180~nm to 230~nm. 

\par
The mechanism of primary scintillation from the second continuum is similar for liquid and gaseous argon. When the incident particle impinges on target atoms, the argon atoms are either excited or ionized. 
During the excitation process, an excited argon atom and two ground-state argon atoms are involved in a three-body collision, creating an excimer (excited dimer Ar$^{*}_{2}$). Alternatively, for the ionization process, an ionized argon atom recombines with electrons and goes through a three-body collision with two ground-state argon atoms to form the excimer. These excimer states are in either singlet ($^{1}\Sigma^{+}_{u}$) or triplet ($^{3}\Sigma^{+}_{u}$) molecular states\cite{0038-5670-26-1-R02}. Subsequently the scintillation light is emitted and the excimer disassociates. The lifetimes for singlet and triplet states are 6 ns (7 ns) and 3.2 $\mu$s (1.6 $\mu$s) in gaseous\cite{doi:10.1063/1.452869} (liquid\cite{PhysRevB.27.5279}) argon, respectively. 

\par
The first continuum shares the same origin as the second continuum but from higher vibrational states\cite{PhysRevA.5.1110}\cite{Goubert1994360}. The origin of the third continuum is still under debate, with several authors postulating different chemical processes \cite{doi:10.1063/1.436447}\cite{PhysRevA.43.6089}\cite{0953-4075-27-9-014}\cite{Grimm1987394}. From the most recent study \cite{Wieser2000233}, at least four different states are involved in the process, possibly including three body collisions of Ar$^{++}$ (Ar$^{+*}$) with ground state argon atoms. This process leads to the creation of Ar$_{2}^{++}$(Ar$_2^{+*}$), which then decays radiatively. 

\par
In general, the relative contribution from each of the three continua depends on the pressure of the argon gas. In particular, the intensity of the first continuum decreases with increasing pressure. At low pressure, the second continuum's contribution is small. The second continuum becomes dominant for pressures larger than approximately 0.8 bar, where the first continuum is negligible~\cite{PhysRevA.5.1110}\cite{Goubert1994360}. The MiniCLEAN detector operates at $\sim$1.5 bar where the second and third continua are observable and contributions from the first continuum are highly suppressed. 

\par
Previous studies have also provided insight on the relative timing of the various scintillation continua. Scintillation light from the third continuum is very fast and contributes mainly to the prompt light~\cite{1748-0221-3-02-P02001}. As previously discussed, light from the second continuum is produced from excimers according to two well-separated time constants: a fast time constant from the singlet state, and a slow time constant from triplet state. According to Ref.\cite{1748-0221-3-02-P02001}, the scintillation light from the second continuum's singlet states is also prompt, but slightly delayed on the order of ten nanoseconds compared to the third continuum component. 

\par
\begin{figure}
\includegraphics[scale=0.3]{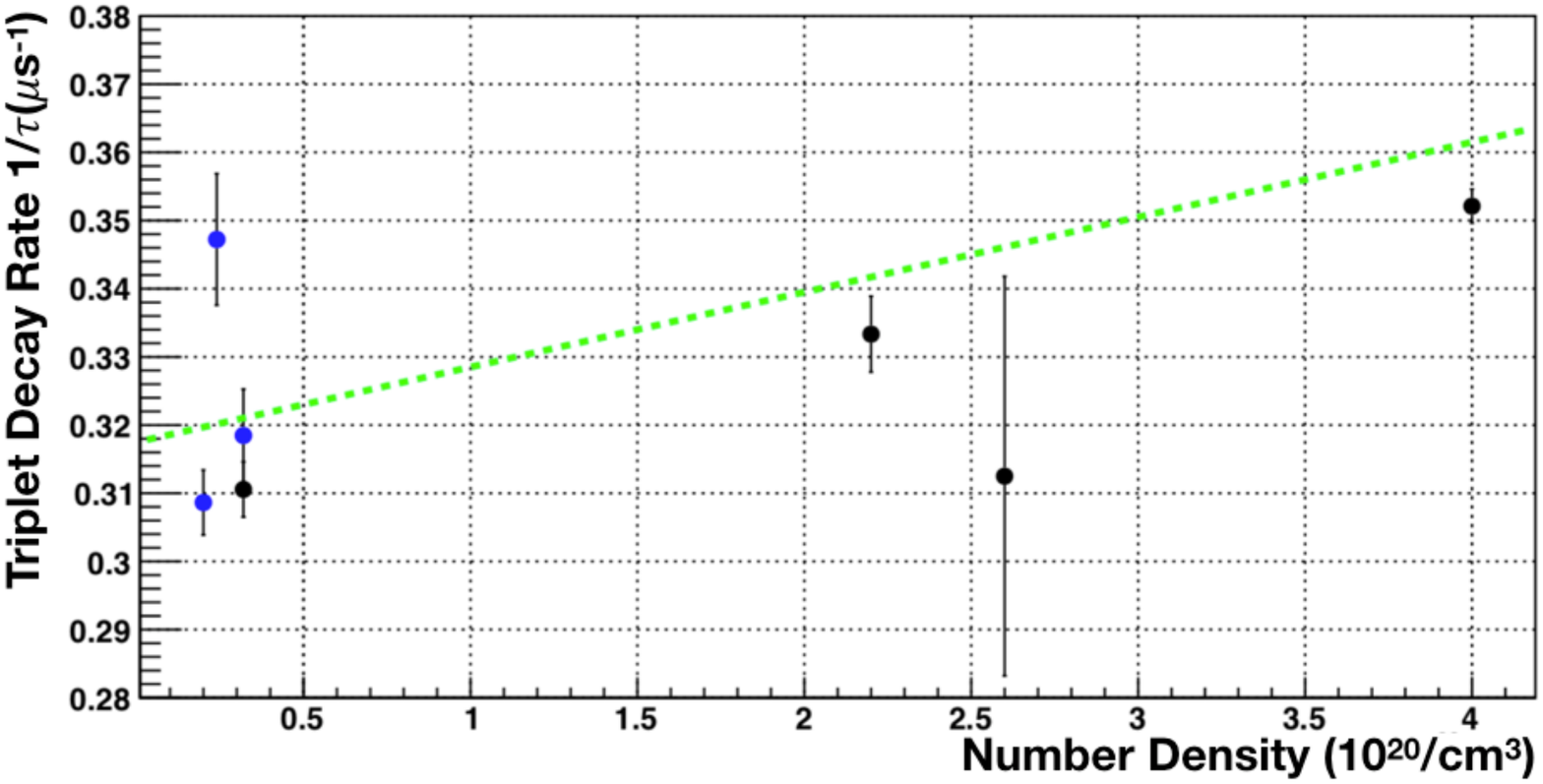}
\caption{\label{fig:ftable} Triplet decay rate (inverse lifetime) versus density from Table ~\ref{tab:table1}.  The green line is a fit of rate versus density from Moutard \cite{doi:10.1063/1.452869}. The blue dots are from more recent results (after 2000) and the black dots are older results.}
\end{figure}

In the presence of impurities (e.g. $\mathrm{O_{2}},\mathrm{N_{2},H_{2}O}$), there is a probability that the argon excimer will undergo a non-radiative collisional reaction with the impurity molecule. This non-radiative quenching process is in competition with the de-excitation process that leads to light emission. The singlet state is not strongly affected by this process, due to its very fast decay time, but with sufficiently high impurity concentrations ($>$ 100 ppm) the singlet state can begin to be quenched as well\cite{1748-0221-6-08-P08003}. However, even impurity concentrations $<100$~ppm can significantly reduce the triplet lifetime and total light yield. As such, the measurement of triplet lifetime can lead to an understanding of argon impurity level. Various triplet lifetime measurements in gaseous argon found in the literature are summarized in Table \ref{tab:table1}. Figure \ref{fig:ftable} plots the same experimental data points along with the function from \cite{doi:10.1063/1.452869} which predicts the behavior of triplet decay rate (inverse lifetime) as a function of density. It can be seen in the plot that the decay rate decreases (or lifetime increases) as density decreases. With a low density, the chances for the dimer state to be formed though three-body collisions is decreased, resulting in smaller decay rates (longer lifetimes) in pure argon. However, even with an impurity level at 1 ppm, the density effect alone can not explain the variation of triplet decay rates and lifetimes. A more complete explanation requires inclusion of the effects of impurities, which have been investigated by many groups\cite{1748-0221-3-02-P02001}\cite{1748-0221-6-08-P08003}\cite{1748-0221-5-06-P06003}\cite{1748-0221-5-05-P05003}. 

\par
In this paper we present the gas triplet lifetime measurements from MiniCLEAN and find the longest triplet lifetime, to our knowledge, that has so far been measured. Section 2 gives an overview of the MiniCLEAN detector and the purification process. Data analysis is presented in detail in Section 3. The results of the analysis of the MiniCLEAN cold gas run are described in Section 4. 
In Section 5, we present the results from the measurement of triplet lifetime and its relationship to the purity. In addition, the quenching effect of each component of light is discussed. In Section 6  we draw conclusions and summarize implications for future detector design.



\begin{table*}
\caption{\label{tab:table1}Triplet lifetime in gaseous argon. The variation of lifetimes is due to both density and (presumably) impurity level.  Only upper limits on impurity are reported. The results are ordered by lifetime. }
\setlength {\tabcolsep}{2pt}
\begin{tabular}{ccccc}
\bf $\tau$ ($\mu$s) & \bf Ref. & \bf Number density ($10^{20}cm^{-3}$) & \bf Estimated impurity level & \bf Induced particle type\\
\hline\hline
2.8 & Thonnard \textit{et al.}\cite{PhysRevA.5.1110} & 0.19 & $<$ 2 ppm&$\beta$\\
2.84$\pm$0.02 & Gleason \textit{et al.}\cite{doi:10.1063/1.434079} & 4 & $<$ 1 ppm&$\beta$\\
2.86 & P. Millet \textit{et al.}\cite{0022-3700-15-17-024} &0.1- 0.26& $<$ 1 ppm&$\alpha$\\
2.88$\pm$0.08&K. Mavrokoridis \textit{et al.}\cite{1748-0221-6-08-P08003} &0.24 & $<$ 1ppb&$\alpha$\\
2.9     &  Carvalho \textit{et al.}\cite{CARVALHO1979487}     &3 & not reported & $\beta$\\
3.0$\pm$0.05& Suemoto \textit{et al.}\cite{Suemoto1977131} &  2.2 & $<$ 10 ppm&$\beta$\\
3.14 $\pm$0.067& C. Amsler  \textit{et al.}\cite{1748-0221-3-02-P02001} & 0.32 & $<$ 9 ppb&$\alpha$\\
3.15$\pm$0.05 & P. Moutard \textit{et al.}\cite{doi:10.1063/1.452869} & 0\footnotemark& $<$ 1 ppm &$ \gamma$ \\
3.2 $\pm$ 0.3& Keto \textit{et al.}\cite{PhysRevLett.33.1365} &  2.6 & $<$ 2 ppm&$\beta$\\
3.22$\pm$0.042& Oka \textit{et al.}\cite{doi:10.1063/1.437923} & 0.32  & $<$ 5 ppm &$\beta$\\
3.24$\pm$0.05&F. Marchal \textit{et al.}\cite{0953-4075-42-1-015201}& 0.2& $<$1 ppm & $\gamma$\\
\hline\hline
\end{tabular}
\footnotetext  1 1. The theoretical natural triplet life time inferred by the data.
\end{table*}



\section{MiniCLEAN experiment}
\label{S:2}
\subsection{Detector overview}
\begin{figure}
\includegraphics[scale=0.25]{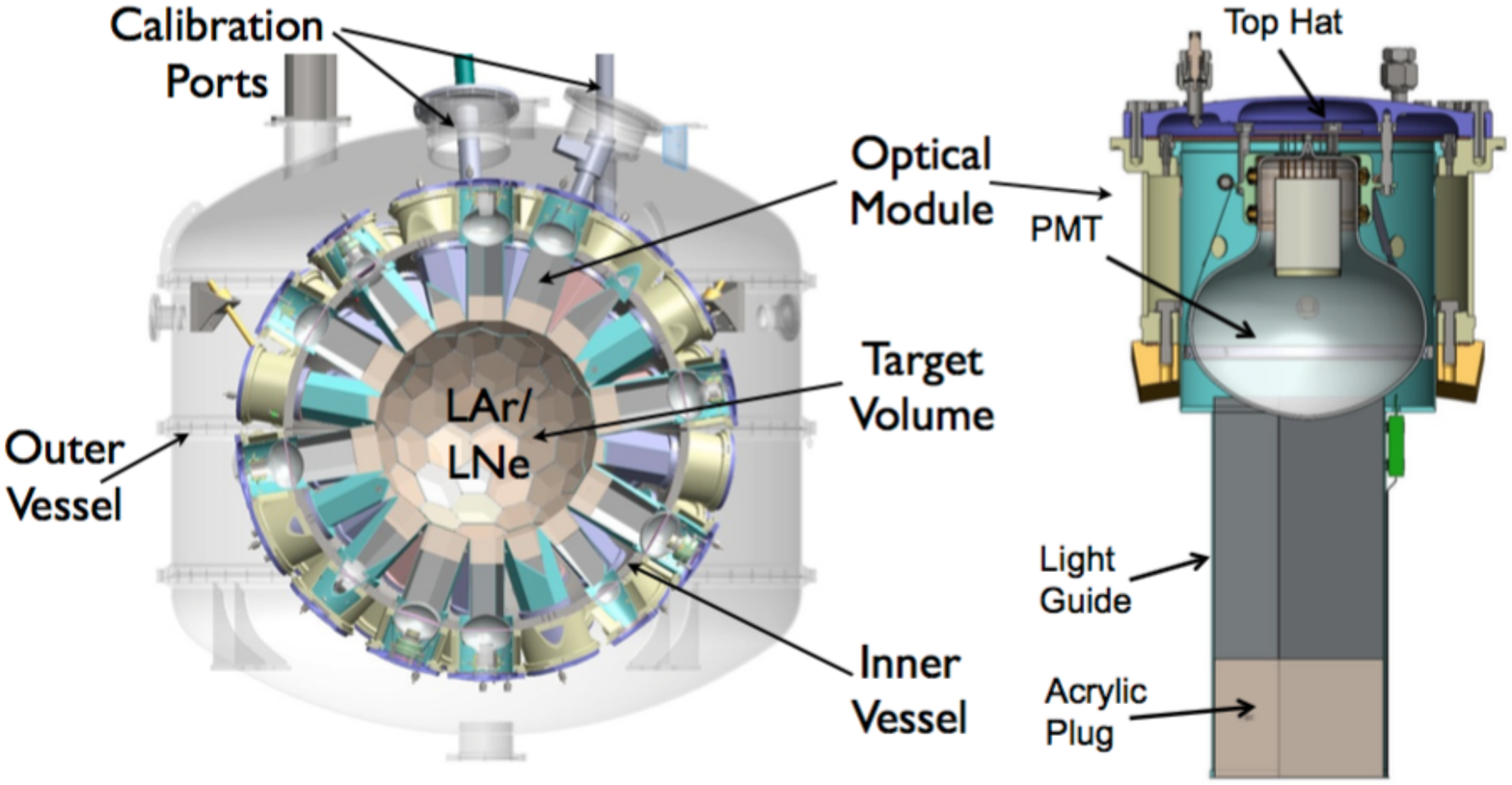}
\caption{\label{fig:fcassette} The MiniCLEAN detector and the optical modules.}
\end{figure}

MiniCLEAN is a single phase liquid argon detector with a 500kg (150 kg) total (fiducial) volume, 
intended for direct dark matter detection. 
The main vessel is placed inside a water tank  to shield the detector from gammas and neutrons from the underground rock. The LAr is held inside a stainless steel Inner Vessel (IV)  which is suspended inside an outer vacuum vessel (OV) to provide thermal insulation. The LAr volume is surrounded by 92  optical cassette modules, arranged spherically  (Fig.~\ref{fig:fcassette}). The optical cassettes consist of acrylic light guides whose inner surface is coated with wave-length shifting tetraphenyl butadiene (TPB) which shifts the argon ultraviolet scintillation light into the visible regime where the PMTs are most sensitive. The light guides are coupled to 8" Hamamatsu R5912-02Mod PMTs.

The supplied voltage to the 92 PMTs is from a VISyN high voltage mainframe and the programmable trip currents can be set for each channel. Both PMT signals and high voltage are carried along a single cable (RG-68 in air, Gore 30 AWG in the OV vacuum) for each PMT. A set of 12 CAEN V1720 250-MHz digitizers receive PMT signals and are configured to obtain 16-$\mu$s waveforms. Each digitizer has eight input channels with 12-bit ADCs and a dynamic range of 2 V. The trigger is defined such that five or more channels are required to cross threshold within a 16-ns coincidence window. A more detailed description can be found in \cite{2017arXiv171102117J}.
\par
The cryogenic system consists of a Gifford-McMahon cryocooler mounted on the OV, with 24 flexible OFHC (Oxygen-Free High thermal Conductivity) copper braids connecting to OFHC copper cold fingers extending into the IV. In addition, a condenser provides supplemental cooling power by liquefying gas as it leaves the purification system. The condenser consists of two concentric cylinders; the outer cylinder vacuum space provides insulation of the inner cylinder, which contains liquid nitrogen. The purified argon gas enters the condenser and passes through a copper coil submerged in the nitrogen, before flowing into the IV.


\subsection{Purification}\label{subsec:clean}
The purification system is responsible for removing the radon and other impurities from the gaseous argon. This prevents potential backgrounds from radon daughters deposited on the TPB, as well as preserving gas purity.  Liquid argon purchased from Air Liquide has purity cited as 99.999~$\%$. This argon is passed through a SAES PS4-MT3-R-1 heated zirconium purifier, to further purify the gas, reducing most impurities (H$_{2}$ , CO, H$_{2}$O, N$_{2}$, O$_{2}$, CH$_{4}$ and CO$_{2}$ in particular) to below ppb level. Subsequently the purified gas passes through an activated charcoal trap, which provides a large surface area kept below the freezing point of radon. The charcoal trap cryo-absorbs radon, while allowing the purified argon to exit the trap. The argon flow is monitored by an RGA (Residual Gas Analyzer), to ensure the purity of the argon gas, before flowing into the IV. \par
During the initial cool down of the IV, an air leak occurred, which caused the triplet lifetime to decrease. In order to remove this impurity from the IV, a series of pump and purge cycles were performed. The RGA was used to monitor the gas impurity level in the IV during the pump and purge cycles. However, this RGA was not sensitive to impurities below 1 ppm, thus the estimated gas exchange during pump and purge cycle was used to infer the impurity level below the RGA limit.  The pump and purge program included 209 cycles, with an average of 6.60 $\pm$ 0.77 \% of the gas pumped out in each cycle. The initial impurity level was 40.86 $\pm$ 2.63 ppm, containing a mixture of mostly oxygen and nitrogen. The impurity level was estimated at the end of each cycle using:
\begin{equation}\label{eq:4}
\frac{dI(t)}{dt} = - \frac{\epsilon f}{T}\cdot I_0.
\end{equation} 
where $I(t)$ is the impurity level as a function of time, $T$ is the pump and purge period, $I_0$ is the initial impurity level, $f$ is the fraction of gas pumped out in each cycle and $\epsilon$ is the contaminant removal (purge) efficiency. 
The variation of temperature was very small ($\ll$ 1K), thus we can ignore the effect from temperature changes during the cycles, and the purge efficiency of the purifier is assumed to be 100~$\%$.\footnote{The getter removes nitrogen and oxygen contaminants to a level of 1ppb.}   We use a recursive form of Eq. (\ref{eq:4}) to estimate the impurity level at the end of each cycle
\begin{equation}\label{eq:5}
I_{i+1} = I_{i}\cdot(1-f_i).
\end{equation}
where $I_i$ and $I_{i+1}$ are the impurity levels for the $i^{th}$ and $(i+1)^{th}$ cycle respectively, and $f_i$ is the fraction of pumped out gas in $i$-th cycle. This gives the impurity level at the end of the full pumping cycle. However, due to the low trigger rates ($\sim$ 8 Hz), we need relatively long data taking times to accumulate events in order to fit the triplet lifetime. We can integrate Eq.(\ref{eq:4}) with respect to time to obtain the average impurity level during the time period T of the pump and purge cycle. 
\begin{equation}\label{eq:6}
I_{Avg} = I_i\cdot\frac{(1-e^{-f_{i+1}})}{f_{i+1}} \approx I_i\cdot\left(1-f_{i+1}/2\right).
\end{equation}
This gives the average impurity level ($I_{Avg}$) for the $(i+1)^{th}$ cycle. In addition, when the gas flow to the IV restores the pressure before the next pump out, the IV is basically at static pressure. Thus the impurity level acquired at the end of the pumping cycle is used to estimate the systematic uncertainties  on impurity level. By comparing the impurity level obtained from two approaches (Eq. (\ref{eq:5}), Eq. (\ref{eq:6})), the estimated systematic uncertainty is $\pm$ 7.1\% for the impurity level inferred from the pump and purge cycle. The systematic uncertainty from the initial value measurement can be estimated using multiple scans right before the beginning of pump and purge cycle. The estimated value is $\pm$ 6.4\%.



\section{Analysis Methods}

\subsection{RAT -- Monte Carlo simulation}
The data analysis is performed using custom software: the RAT analysis tool\cite{rat}. RAT was originally developed by the Braidwood collaboration\cite{BOLTON2005166}, and is currently used by MiniCLEAN\cite{RIELAGE2015144}, DEAP-3600\cite{PhysRevLett.121.071801}, and SNO+\cite{CHEN200565}. RAT incorporates the electromagnetic and hadronic physics simulation methods provided by GEANT4\cite{2006ITNS}\cite{Agostinelli2003250}, the data storage and processing tools provided by ROOT\cite{rootu}, parts of the scintillation and PMT simulation from GLG4sim (an open source package released by the KamLAND collaboration)\footnote{Generic Liquid Scintillator GEANT4 simulation, written and maintained by Glenn Horton-Smith from the KamLAND collaboration.} and much of the design philosophy of the SNOMAN tool developed by the SNO collaboration\cite{snoman}. The aim of RAT is to provide an accessible and easily configurable framework for simulation and analysis in scintillator/PMT based experiments and test stands.
\subsection{Reconstruction}
\begin{figure}
\includegraphics[scale=0.3]{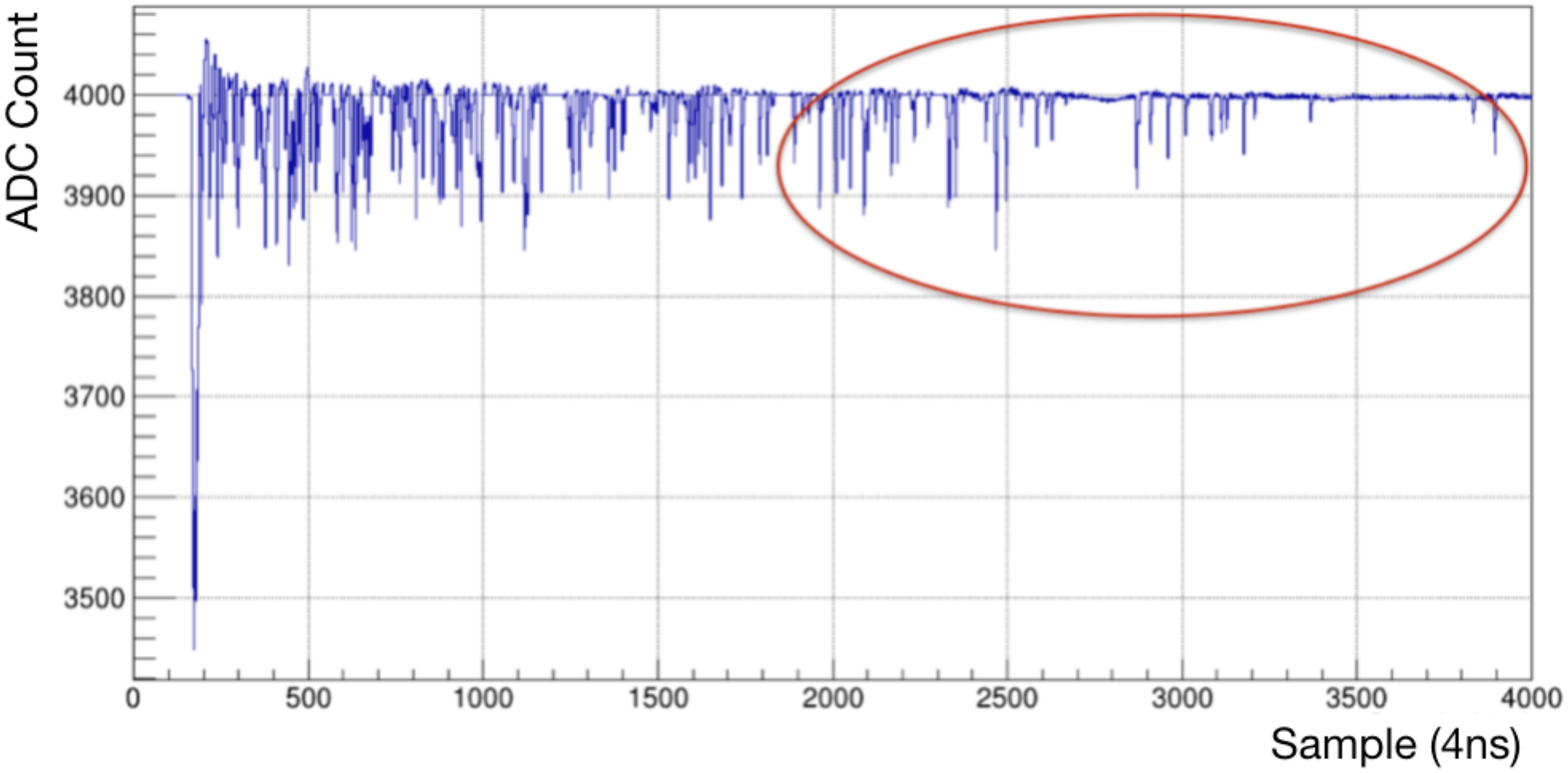}
\caption{\label{fig:fexampleraw} Raw waveform of a scintillation event. The region circled is dominated by single photon pulses.}
\end{figure}

\begin{figure}
\includegraphics[scale=0.26]{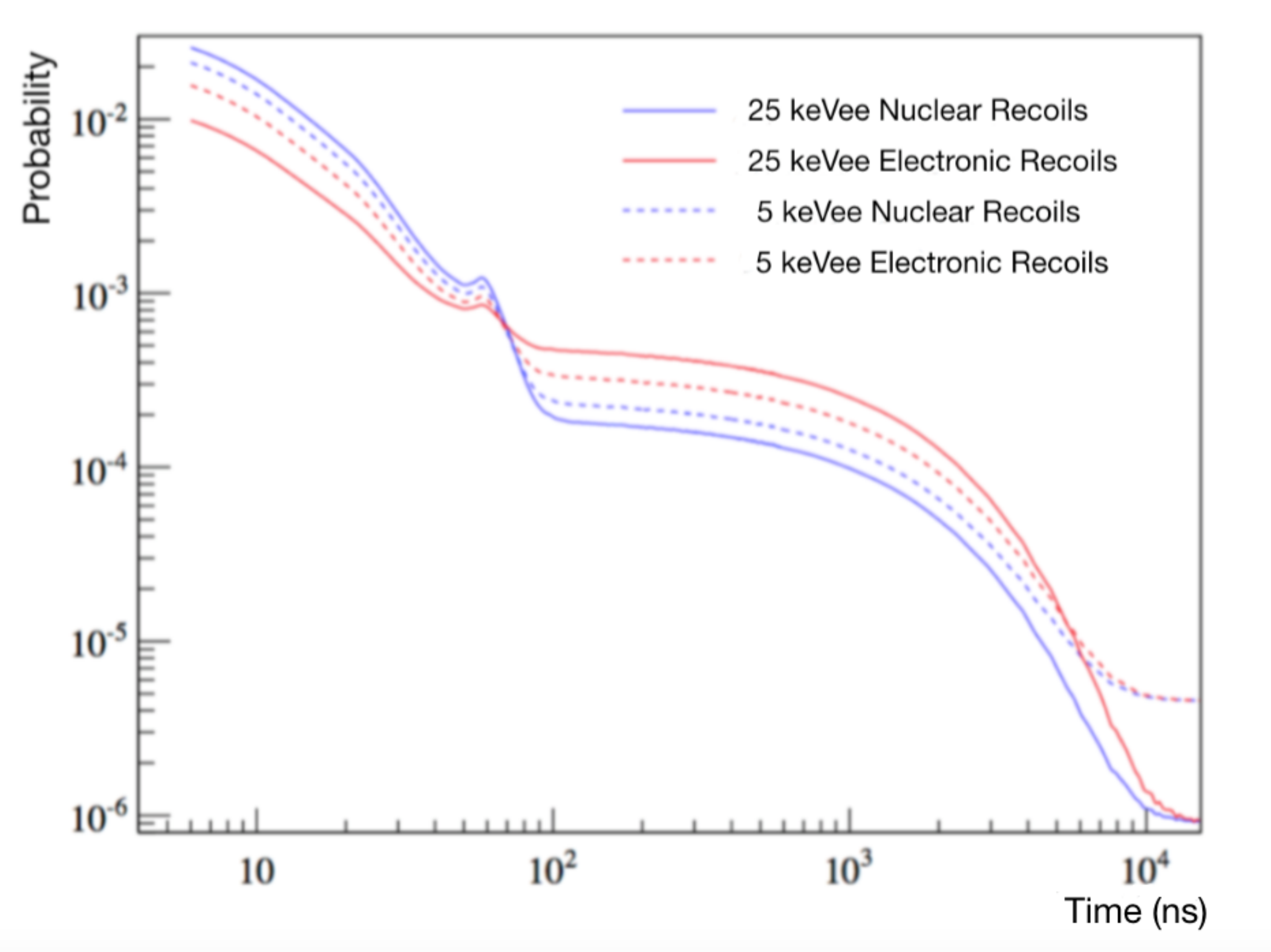}
\caption{\label{fig:fscin} Photoelectron detection time PDFs for electronic and nuclear recoils at 5 keV$_{ee}$ and 25 keV$_{ee}$ energies from MiniCLEAN Monte Carlo simulation. The bump around 60 ns is from PMT late pulsing.}
\end{figure}

\begin{figure}
\includegraphics[scale=0.45]{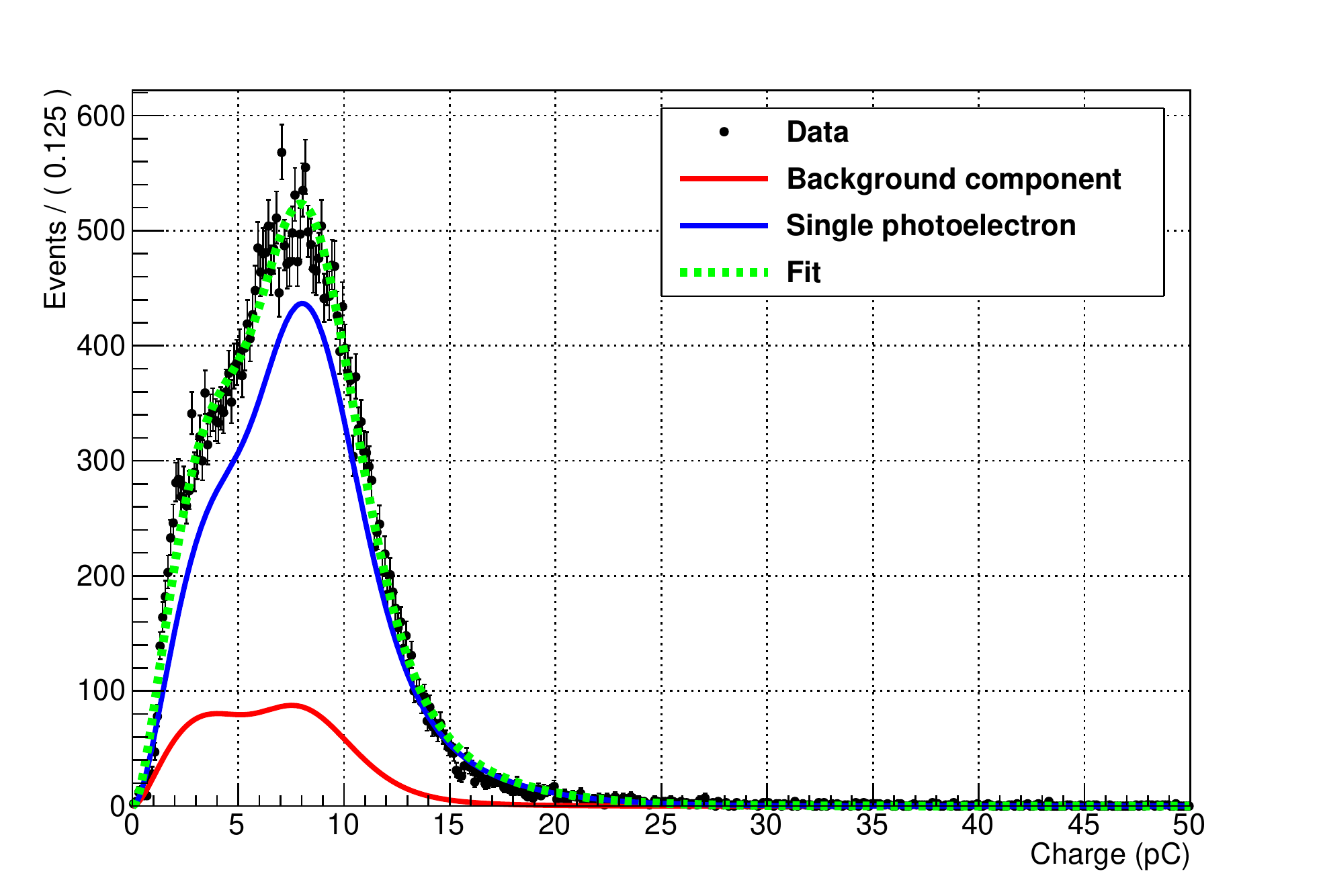}
\caption{\label{fig:fexampleSPEfit} Example of single photoelectron charge distribution fitted with double gamma distribution. Red curve represent the background component , blue curve is the contribution from single photoelectrons and the green dashed line is from the fit. The estimated SPE value from the fit for this distribution is 7.96 pC.}
\end{figure}

PMTs are calibrated using late scintillation light, which is of low intensity and dominated by single photons (see Fig.  \ref{fig:fexampleraw}). 
Using the timing probability density function (PDF) from Fig. \ref{fig:fscin} \cite{AkashiRonquest201540}, we calculate the pile-up probability and identify the part of the event dominated by single photons, which is subsequently used to calculate SPE charge distributions for each PMT. The SPE charge distribution contains two components: the contribution from real single photons, and the contribution from various sources of backgrounds including both dark hits and electronic noise. The background component 
is included in the fit, using a distribution obtained by a periodic trigger. The SPE distribution is fitted with a double-gamma function\cite{tomPMT} as shown in 
Fig. \ref{fig:fexampleSPEfit}.\par
The position reconstruction is performed by calculating the charge centroid for each event, defined as: 
\begin{equation}
\vec{R} = \frac{\Sigma_i \vec{r}_i\cdot Q_i^2}{\Sigma_iQ_i^2}.
\end{equation}
where $\vec{r}_i$ is the position of the $i^{th}$ PMT and $Q_i$ is the charge in that PMT.


\subsection{Discrimination parameters}
In order to select electronic recoil events, two discrimination parameters were defined : $F_{\rm prompt}$ ($F_{\rm p}$) and Charge Ratio ($Q_R$). 
  $F_{\rm prompt}$ is the fraction of hits occurring in a prompt window, and is defined as: 
\begin{equation}\label{eq:1}
F_{\rm p} = \frac{\int^{\epsilon}_{T_{i}} V(t) dt}{\int^{T_{f}}_{T_{i}} V(t) dt}.
\end{equation} 
where $V(t)$ is the ADC counts of the raw waveform, $T_{i}$ is the event start time before the prompt peak, $T_{f}$ is the end time of the event window and $\epsilon$ is an intermediate time, chosen depending upon the timing characteristics of the scintillator. $F_{\rm prompt}$ is a useful discriminant since scintillation events created by different particle types have very different timing profiles, as seen in Fig. \ref{fig:fscin}. 
 $F_{\rm prompt}$ is used to discriminate between electronic and nuclear recoils. In MiniCLEAN, $T_{i}$ is set to be $-28$~ns, $\epsilon$ is 80~ns, $T_{f}$ is 1.5~$\mu$s and the total data acquisition window is 16~$\mu$s.\par 
The total charge is defined as the sum of charges measured in all PMTs, and the charge ratio (Q$_R$) is defined as the maximum charge registered on any individual PMT divided by the total charge of the event. For example, Q$_R$ will be equal to one if only a single PMT registers the full charge in a given event. Q$_R$ is thus an indirect measure of the isotropy of recorded PMT hits in an event.


\subsection{Pulse Finding}
\label{pulseFinding}
\begin{figure}
\includegraphics[scale=0.32]{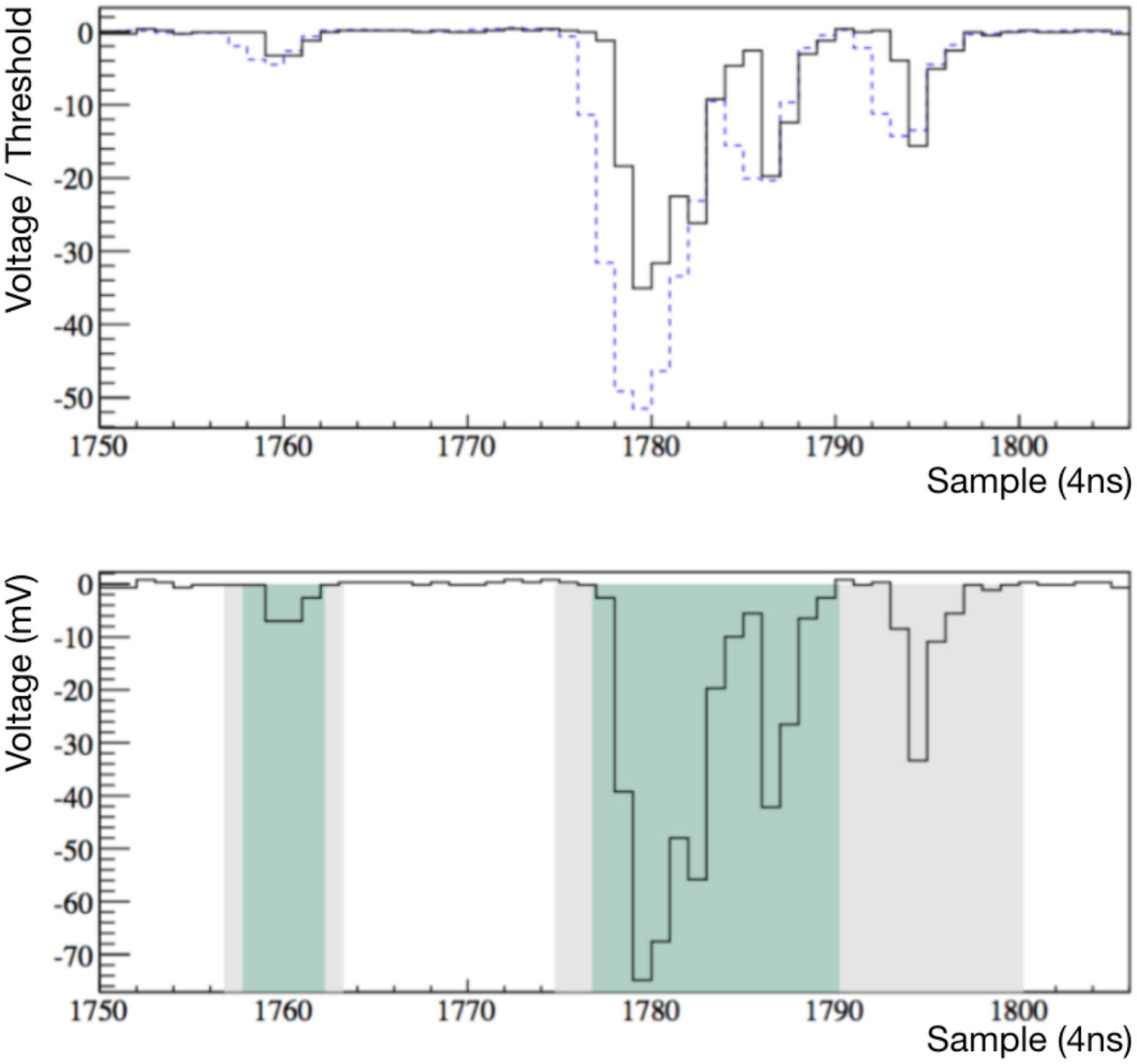}
\caption{\label{fig:fpulsefinding} A typical voltage waveform from a single PMT in the MiniCLEAN Monte Carlo simulation. The top panel shows the waveform normalized by 5 times the RMS of the electronic noise profile (black, solid) compared to the sliding integral value normalized by the corresponding threshold (blue, dashed). The sliding integration window enhances the right-skew PMT pulses relative to a threshold, while providing a filter for high frequency electronics noise. The bottom panel shows the pulse regions identified by the pulse finder. Green shaded regions are the regions where threshold is crossed, and the gray regions indicate a buffer region that extends the pulse boundaries. If threshold is crossed again within the buffer, the pulse boundary is further extended as in the right-most pulse region.}
\end{figure}

Calibrated waveforms are scanned for each PMT separately with a sliding 12-ns (3-sample) integration window, and a pulse region is defined whenever the integral exceeds 5 times the RMS of noise samples multiplied by the square root of the number of samples in the window. Once the pulse threshold has been crossed, the end boundary of the pulse region is defined as the time when the sliding window integral drops below the RMS divided by the square root of the number of samples. Figure \ref{fig:fpulsefinding} shows the sliding window calculation applied to a simulated pulse created by 5 photoelectrons.


\begin{figure*}
\centering
\begin{subfigure}[c]{.4\linewidth}
    \hbox{\hspace{-3em} \includegraphics[width=9cm]{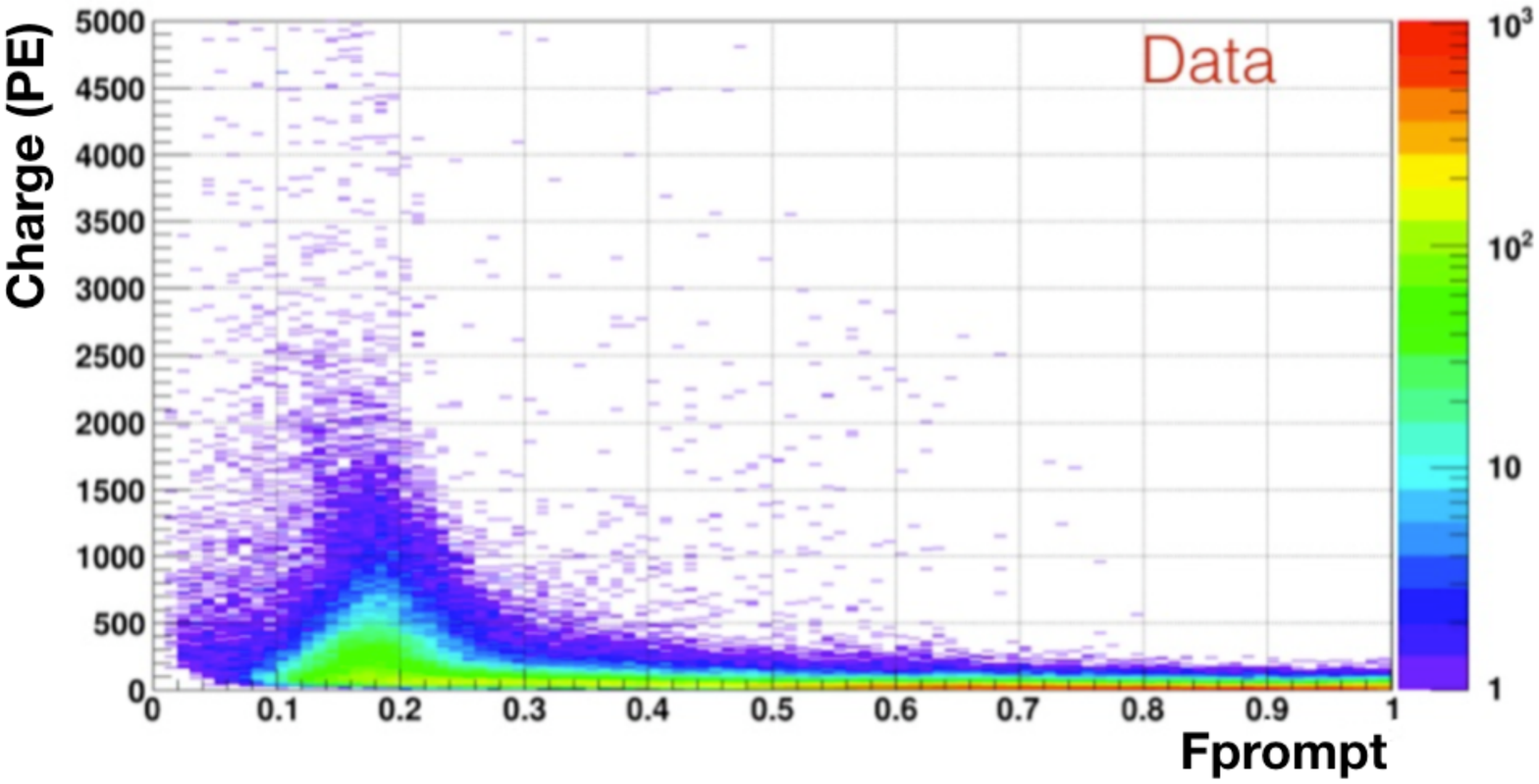}}
    \caption{}
    \label{fig:fp_q_old}
\end{subfigure}\hspace{12pt}
\begin{subfigure}[c]{.4\linewidth}
    \centering
    \includegraphics[width=9cm]{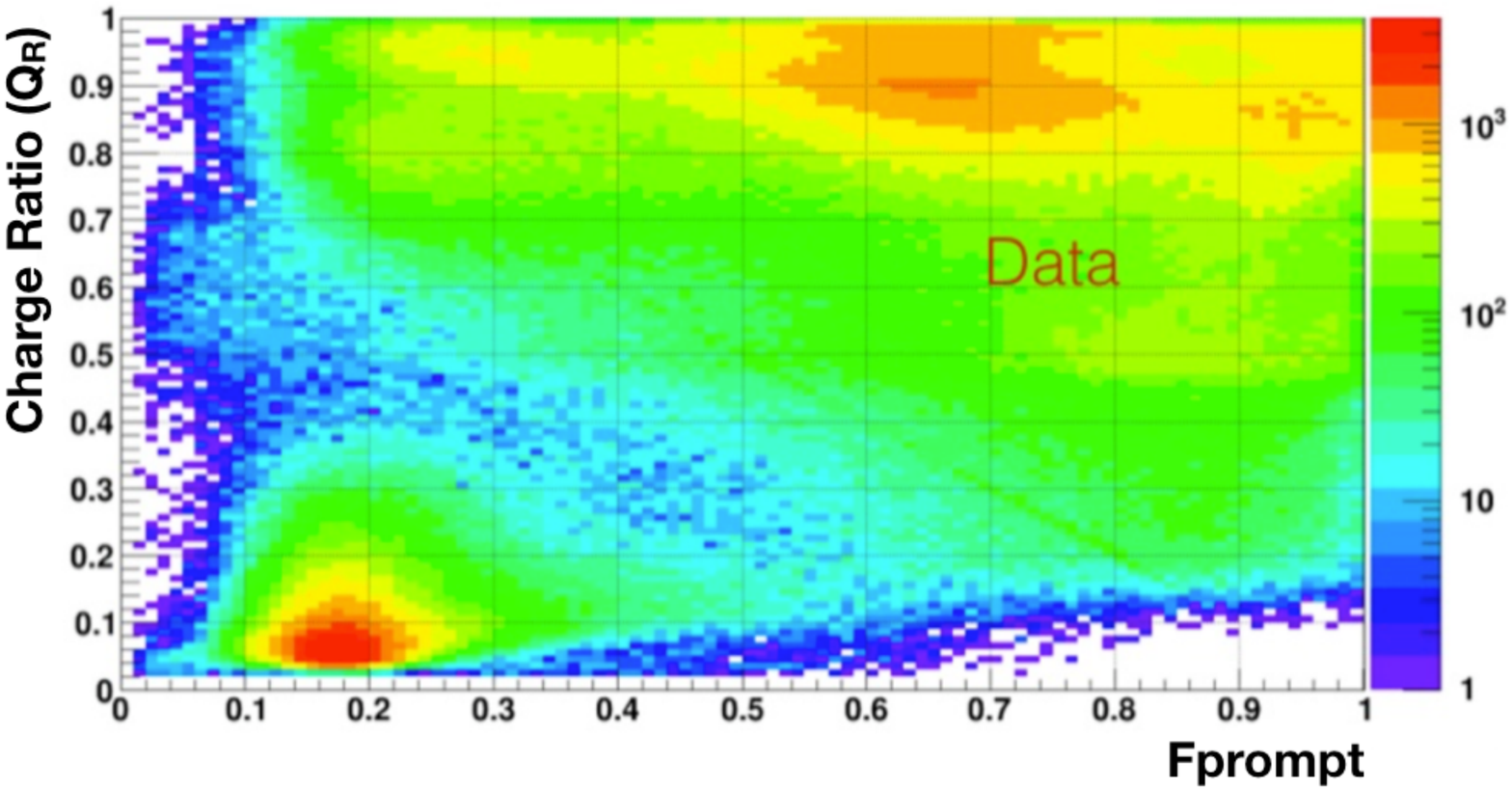}
    \caption{}
    \label{fig:fp_qratio_old}
\end{subfigure}\vspace{12pt}
\begin{subfigure}[c]{.4\linewidth}
    \centering
    \hbox{\hspace{-1em}\includegraphics[height=6cm]{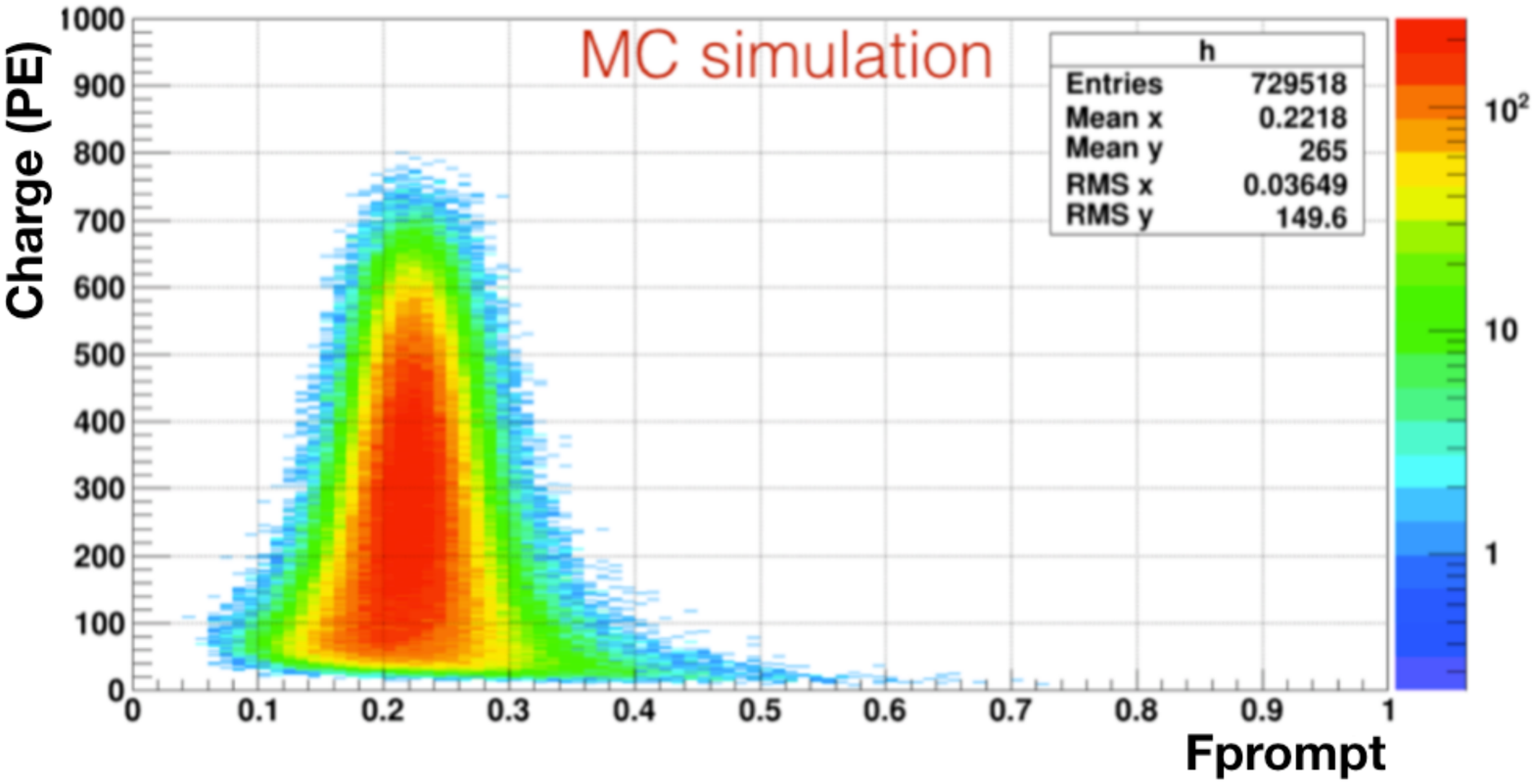}}
    \caption{}
    \label{fig:Charge_fp_MC}
\end{subfigure}\hspace{24pt}
\begin{subfigure}[c]{.4\linewidth}
    \centering
    \includegraphics[height=6cm]{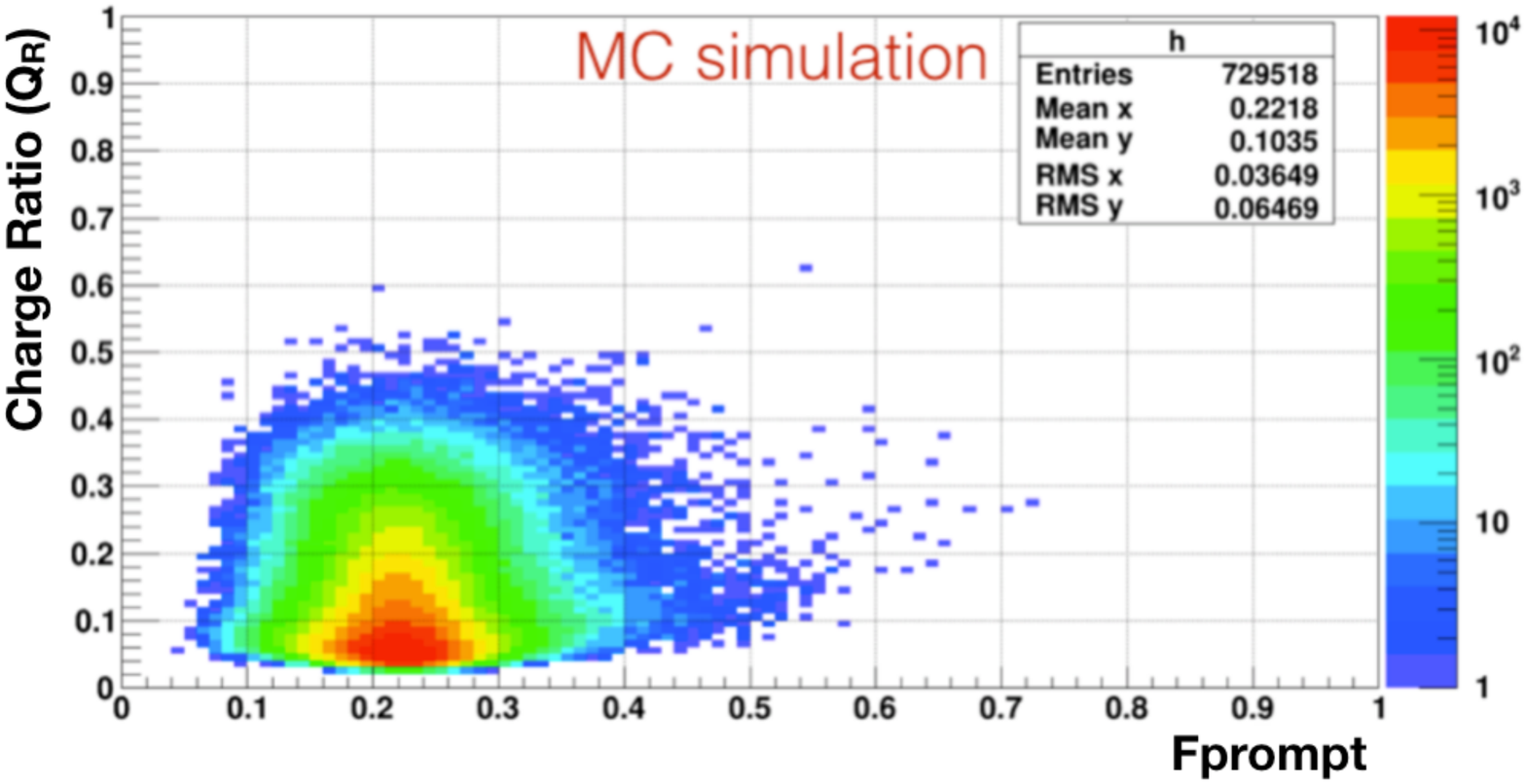}
    \caption{}
    \label{fig:Qratio_fp_MC}
\end{subfigure}
\caption{(a) Charge vs $F_{\rm prompt}$  from the data. The peak in the low $F_{\rm prompt}$ region is from electronic recoils in gaseous argon from the data. (b) Charge Ratio (Q$_R$ ) vs $F_{\rm prompt}$  from the data. The group of events at low Q$_R$  and low $F_{\rm prompt}$ is from electronic recoils which correspond to the peak in the Charge-$F_{\rm prompt}$ plot. (c) Charge vs $F_{\rm prompt}$ from Monte Carlo simulation of $^{39}$Ar events. Note the scale on the y-axis is different from (a). (d) Charge Ratio(Q$_R$ ) vs $F_{\rm prompt}$ from Monte Carlo simulation of $^{39}$Ar events.}
\label{fig:fcharge}
\end{figure*}



\begin{figure*}
\centering
\begin{subfigure}[c]{.4\linewidth}
    \hbox{\hspace{-3em} \includegraphics[width=9cm]{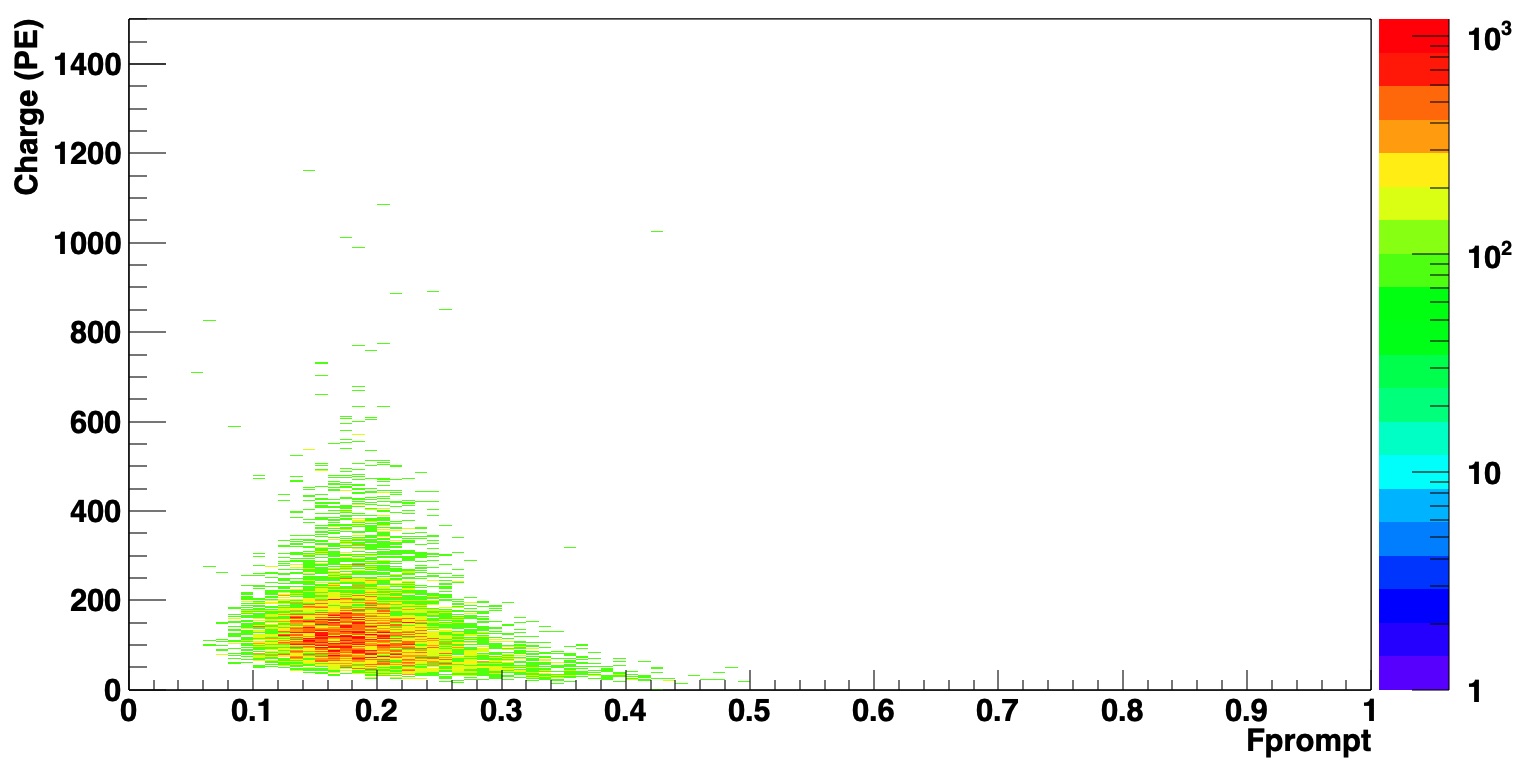}}
    \caption{}
    \label{fig:fp_q_old}
\end{subfigure}\hspace{12pt}
\begin{subfigure}[c]{.4\linewidth}
    \centering
    \includegraphics[width=9cm]{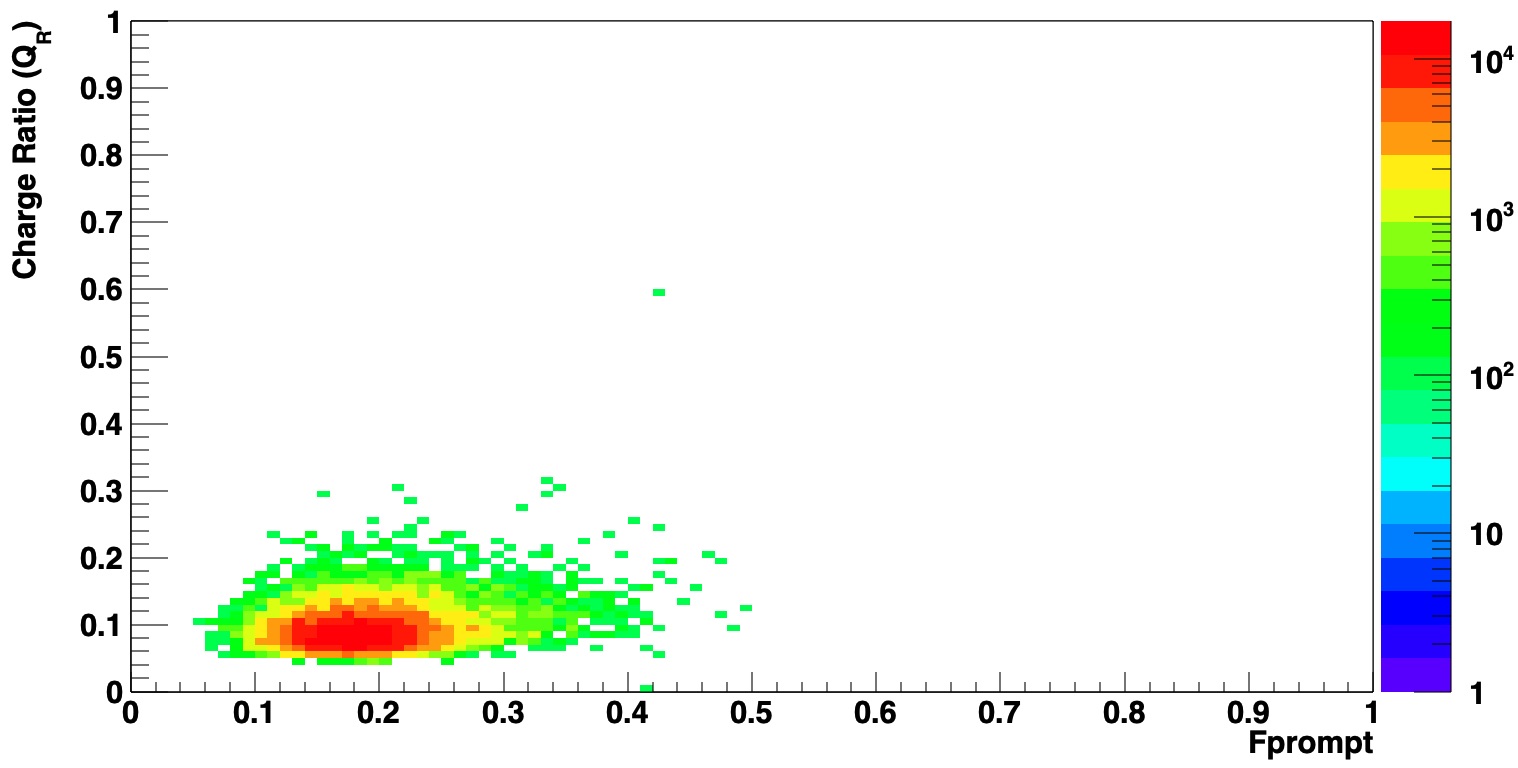}
    \caption{}
    \label{fig:fp_qratio_old}
\end{subfigure}\vspace{12pt}
\caption{(a) Charge vs $F_{\rm prompt}$  after applying all cuts. (b) Charge Ratio (Q$_R$ ) vs $F_{\rm prompt}$  after applying all cuts.}
\label{fig:fchargecut}
\end{figure*}


\section{Data}
\subsection{Data sets}
Data used in this analysis were taken from February to April 2017, with a total of 39 runs and 18.4 days of live time. Initially 37 of the 92 PMTs were excluded from this analysis due to various reasons (no connection, low gain, high noise rates.) The number of total well-functioning PMTs included in the final analysis are 62 (55 for early pump and purge runs, 7 more PMTs were recovered in the later runs). The data taking started approximately 5 hours after the beginning of the pump and purge cycle; the initial motivation was to monitor the cycle progress. After completion of the pump and purge cycle, IV was kept static to observe the stability of the triplet lifetime.

\subsection{Event selection}
The dominant scintillation events are from intrinsic $^{39}$Ar beta decays. In this analysis, the scintillation events were chosen by selecting events with low $F_{\rm prompt}$ and low Q$_R$. These electronic recoil scintillation events have low $F_{\rm prompt}$ value because they contain more late light than nuclear recoil like events (Fig. \ref{fig:fscin}). These events have a small Q$_R$ because scintillation light in gaseous argon is emitted isotropically, thus multiple PMTs are hit.
In addition, the high $Q_{R}$ regions is dominate by instrumental events (i.e. cross talk between channels) and the Cherenkov radiation in acrylic produced by residual radioactive in PMT glass or high energy external gamma is dominate in high $F_{prompt}$ region.
 These event selections were validated using a Monte Carlo simulation (with the same number of PMTs turned on), as shown in Fig. \ref{fig:fcharge} (c) and (d). Event selection values were defined as $F_{\rm prompt}<$0.5 and Q$_R$ $<$0.6. The cut efficiency estimated using MC simulation is 99.988$\%$.  We noted that the peak of  $F_{\rm prompt}$ in simulation is at smaller value than in the data. We believe that this discrepancy is due to the fact that in the simulation we use the average triplet lifetime value (acquired from the data) whereas in the data we use the individual PMT lifetimes which have a slight variation possibly due to the different efficiency for PMT in different locations. \par
\begin{figure}[!htb]
\includegraphics[scale=0.14]{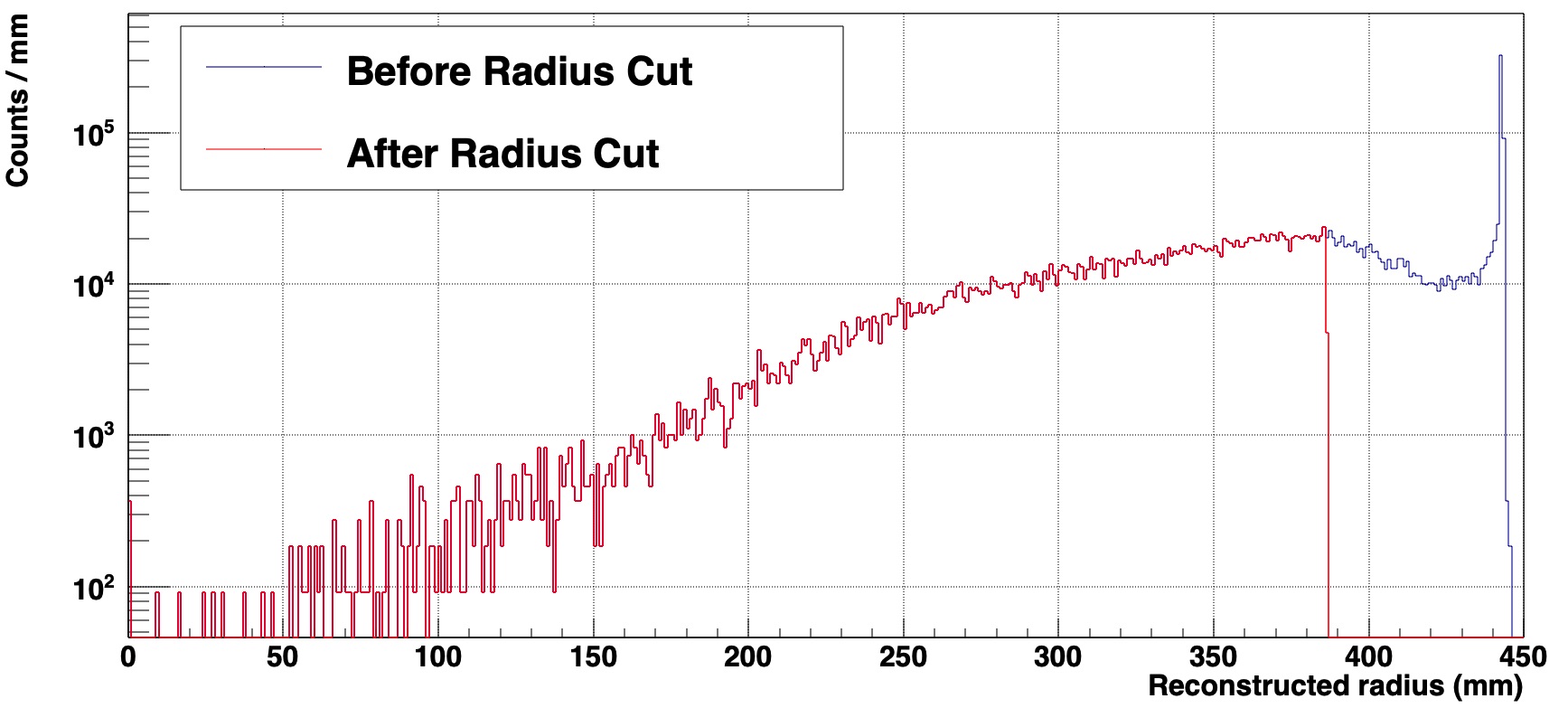}
\caption{\label{fig:fnumber} Number of counts vs centroid radius (mm).  }
\end{figure}

PMT base discharge is a problem when operating in gaseous argon, where the PMT bias voltage is sufficient to cause breakdown over distances less than a centimeter. Most bases were coated in epoxy, so the breakdown rate in gas is relatively low and bases which were not conformally coated remained off during the cold gas runs. However, once a base begins to discharge, the spark path provides a preferred route for continued discharge. Figure \ref{fig:fnumber} shows the reconstructed radius of each event passing the $F_{\rm prompt}$ and Q$_R$  cut using a charge centroid $\vec{R}$. The existence of PMT discharge events in one channel biases the centroid reconstructed radius toward the edge of the detector, resulting in excessive events reconstructed near the edge of the IV. Therefore, to remove these discharge events an additional radius cut was applied to the data and defined as $(R/R_{TPB})^3 < 0.7$, where R is the centroid reconstructed radius, and $R_{TPB}$ is the radius of the TPB (435 mm). The cut efficiency determined by MC simulation changed to 88.30$\%$. The results are shown in Fig. \ref{fig:fchargecut}\par

An additional type of event is frequently observed, one in which only PMTs on
the same WFD board have charge, corresponding to PMTs on one side of the
detector.  
Such events arise from electronic cross-talk on the WFD board.  We removed
these events by a dedicated  cut.  


\section{Results}
\subsection{Triplet lifetime}
\begin{figure}
\includegraphics[scale=0.3]{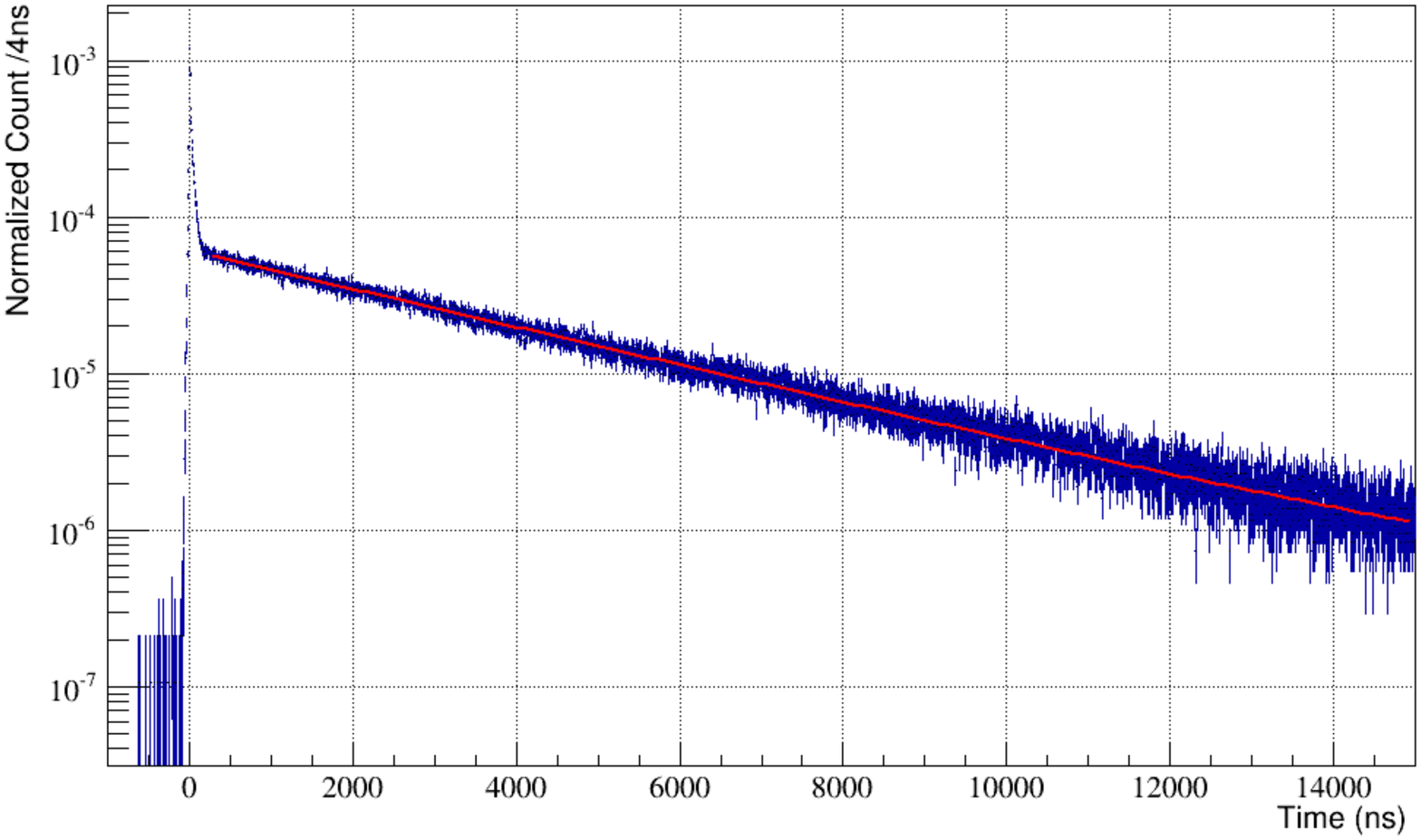}
\caption{\label{fig:ftrip1}   Pulse time distribution fit with fit function from Eq. (\ref{eq:2}).  }
\end{figure}

The start times found by the pulse-finding algorithm (Section~\ref{pulseFinding})
on the selected data is fitted with a simple exponential function plus constant background model:
\begin{equation}\label{eq:2}
F(t) = p_0 \cdot[ (1-p_1) e^{ -t/\tau } +p_1].
\end{equation}
where $p_{0}$ is a normalization, $p_{1}$ is the background fraction and $\tau$
is the triplet lifetime. The window for the fit is taken from 200 ns after time
zero to the end of the data acquisition window, to prevent contamination from
the scintillation light of the singlet state. A sample fit is shown in Fig
\ref{fig:ftrip1}. \par


\begin{figure}
\includegraphics[scale=0.48]{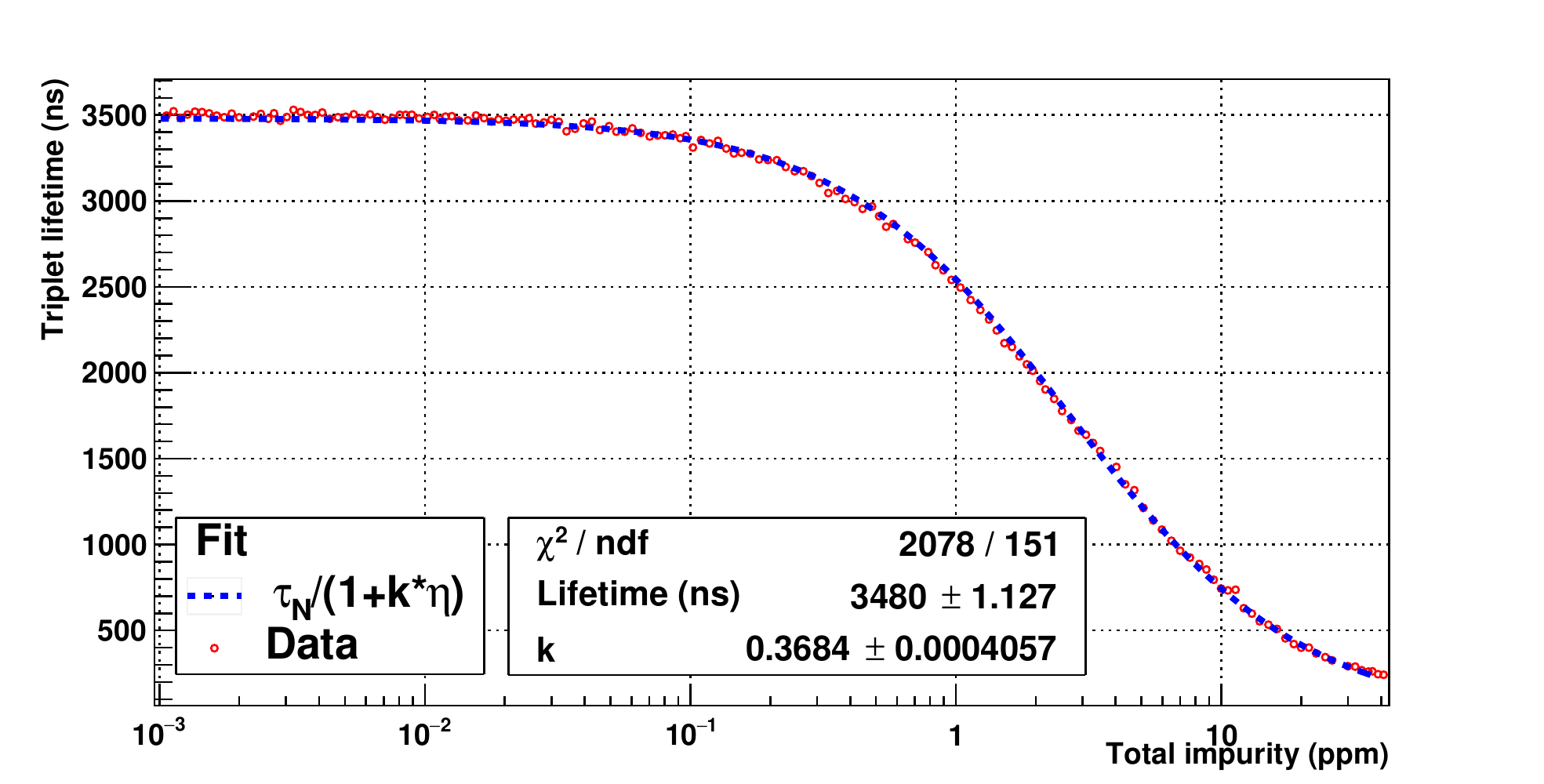}
\caption{\label{fig:fimpurity1}   Triplet lifetime vs total impurity level. The blue dashed line is the fitted function (Eq. \ref{eq:7}). }
\end{figure}

The method to assess impurity level was described in Sec. \ref{subsec:clean}. The pulse time distribution for each pump and purge cycle is fit to get the triplet lifetime for each cycle. The triplet lifetime and the associated impurity level is shown in Fig. \ref{fig:fimpurity1}. The fitting function of the curve is :
\begin{equation}\label{eq:7}
\tau_{m} = \frac{\tau_{N}}{1+ k\cdot\eta }.
\end{equation}
where $\tau_{m}$ is the measured triplet lifetime, $\tau_{N}$ is a fit parameter representing the natural triplet lifetime with zero impurity present in gaseous argon, $\eta$ is the total impurity level in ppm and $k$ is the fitting constant. \par
\begin{figure}
\includegraphics[scale=0.15]{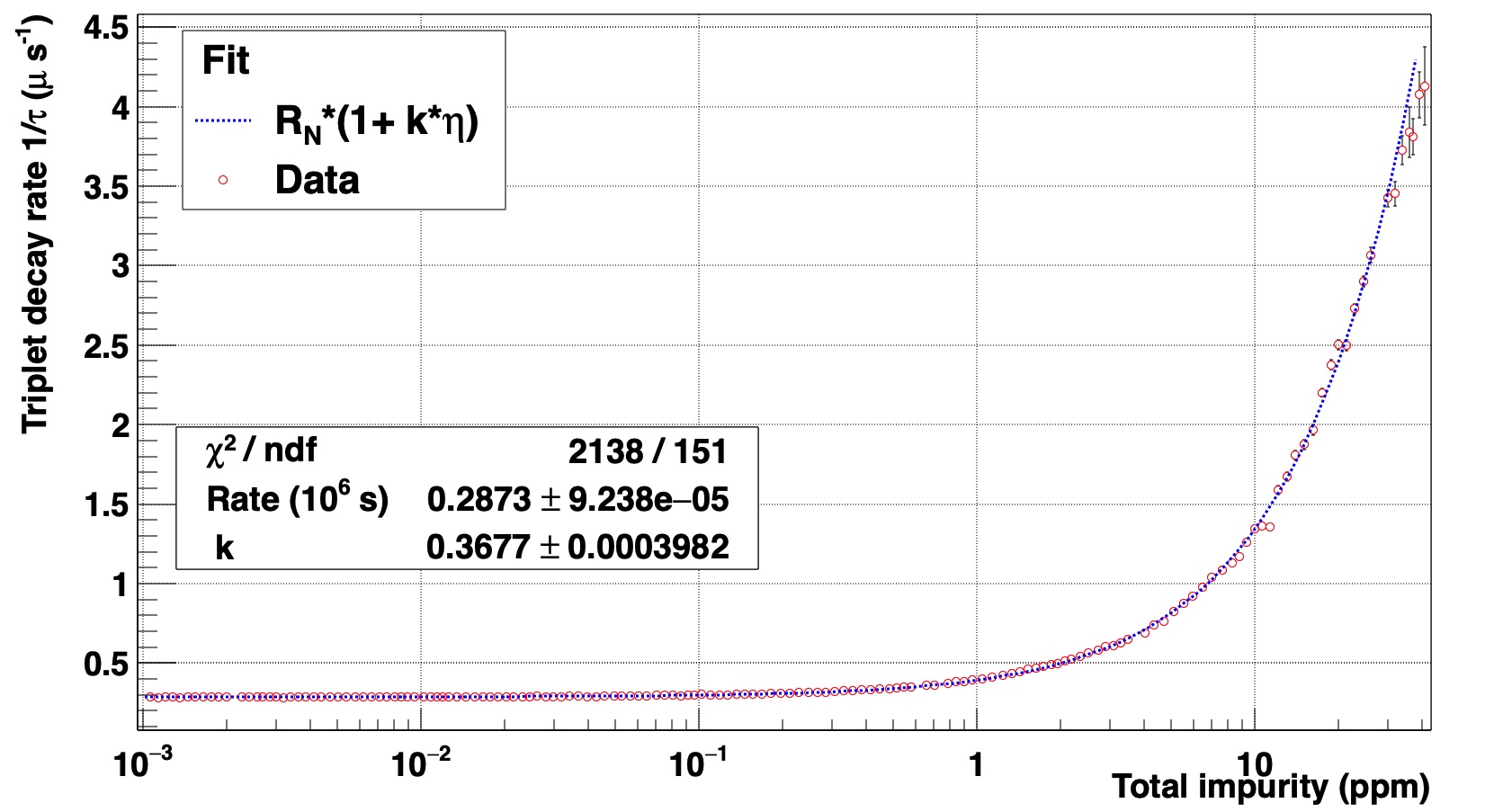}
\caption{\label{fig:fimpurity2}   Triplet decay rate vs total impurity level. The blue dashed line is the fitted function (Eq. \ref{eq:17}).  }
\end{figure}

Alternatively, we can convert the lifetime to the decay rate and use the inverse function of Eq. \ref{eq:7} to fit a line (Fig. \ref{fig:fimpurity2})
\begin{equation}\label{eq:17}
R_{m} = R_N\cdot (1+ k\cdot\eta).
\end{equation}
where $R_m$ and $R_N$ are the inverse lifetimes (decay rates).
The product of $R_N$ and $k$ is the reaction rate per ppm between argon and the impurity molecule. The reaction rate between argon and impurity molecules has been measured by various groups\cite{doi:10.1063/1.436447}\cite{doi:10.1063/1.437923}\cite{doi:10.1063/1.441532}. The dominant impurity species in MiniCLEAN are oxygen and nitrogen at
the operating temperatures ($<$140 K). The quenching effects from nitrogen diminish with impurity level less than 1 ppm\cite{1748-0221-6-08-P08003}, therefore we assume the quenched light yield is mainly due to oxygen when the impurity 
level is below 1 ppm.

We note that the $\chi^{2}$ of both fits are large, due to a discrepancy in the high impurity level region. This implies 
either the functional form does not describe the behavior well at high impurity levels ($>$ 1ppm) or some additional systematic uncertainty occurs in that region. We know of no systematic effect in our analysis that could account for the difference.  We speculate that the functional form is altered at large impurities due to nitrogen which is known to only significantly quench the triplet state for concentrations above 1 ppm\cite{1748-0221-6-08-P08003}.


\subsection{Component Light yield}
The charge distribution for different scintillation light components was obtained using the SPE calibrated charge.  A charge cut (charge $>$ 0.2 PE) was applied to eliminate background from electronic noise. 
A new SPE counting algorithm\cite{AkashiRonquest201540} is adopted to improve the precision on determining the photon arrival time.
 This algorithm exploits features in the scintillation time structure as prior inputs of a Bayesian probability distribution to estimate the plausible arrival time of the single photoelectrons. 

\begin{figure}
\includegraphics[scale=0.29]{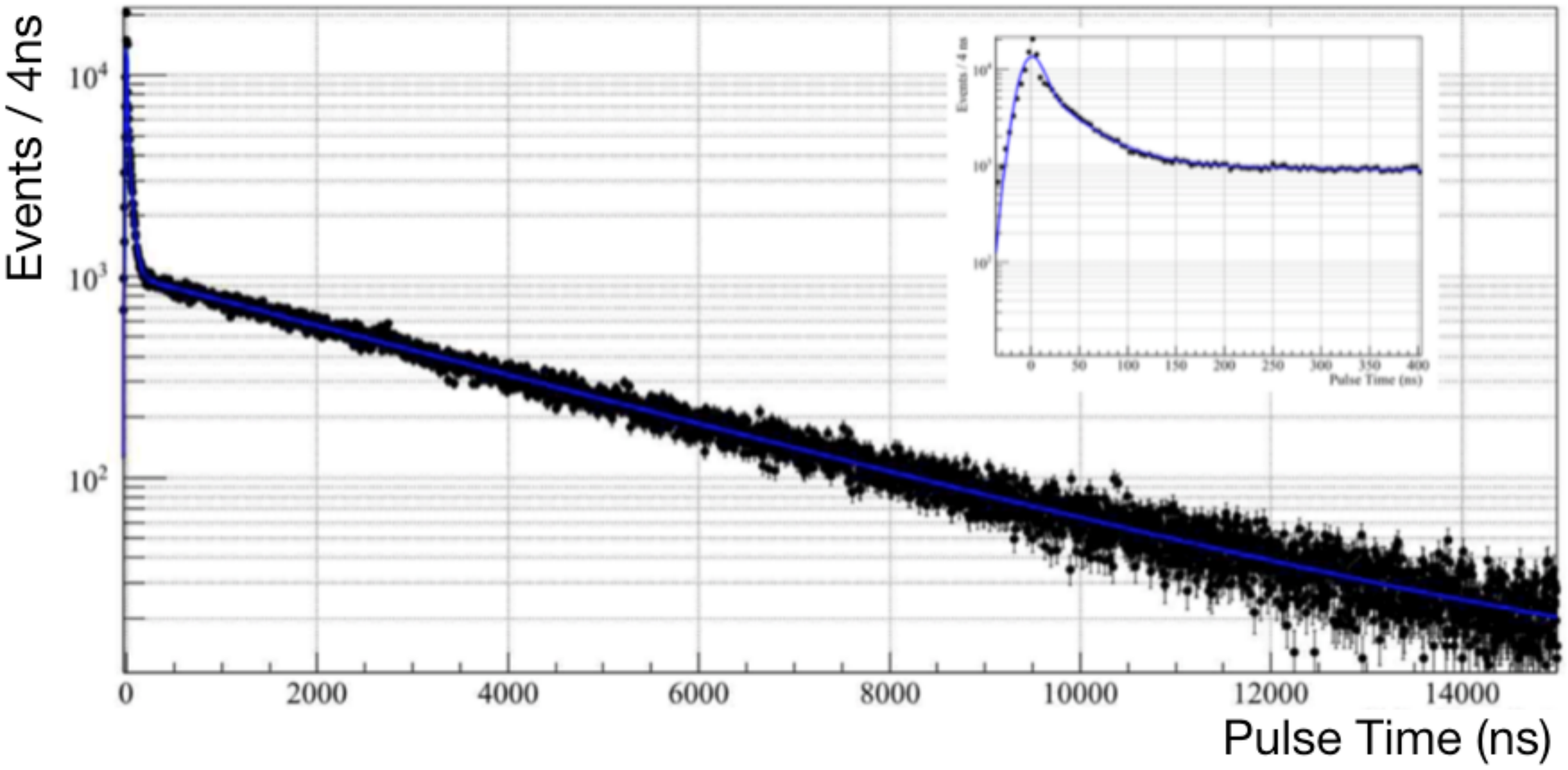}
\caption{\label{fig:ftrip2} Example fit of single photoelectron arrival time from scintillation events with three exponentials convolved with a gaussian resolution function (Eq. (\ref{eq:3})) in cold gas. The inset plot shows this data during the first 400 ns. }
\end{figure}
\begin{figure}
\includegraphics[scale=0.125]{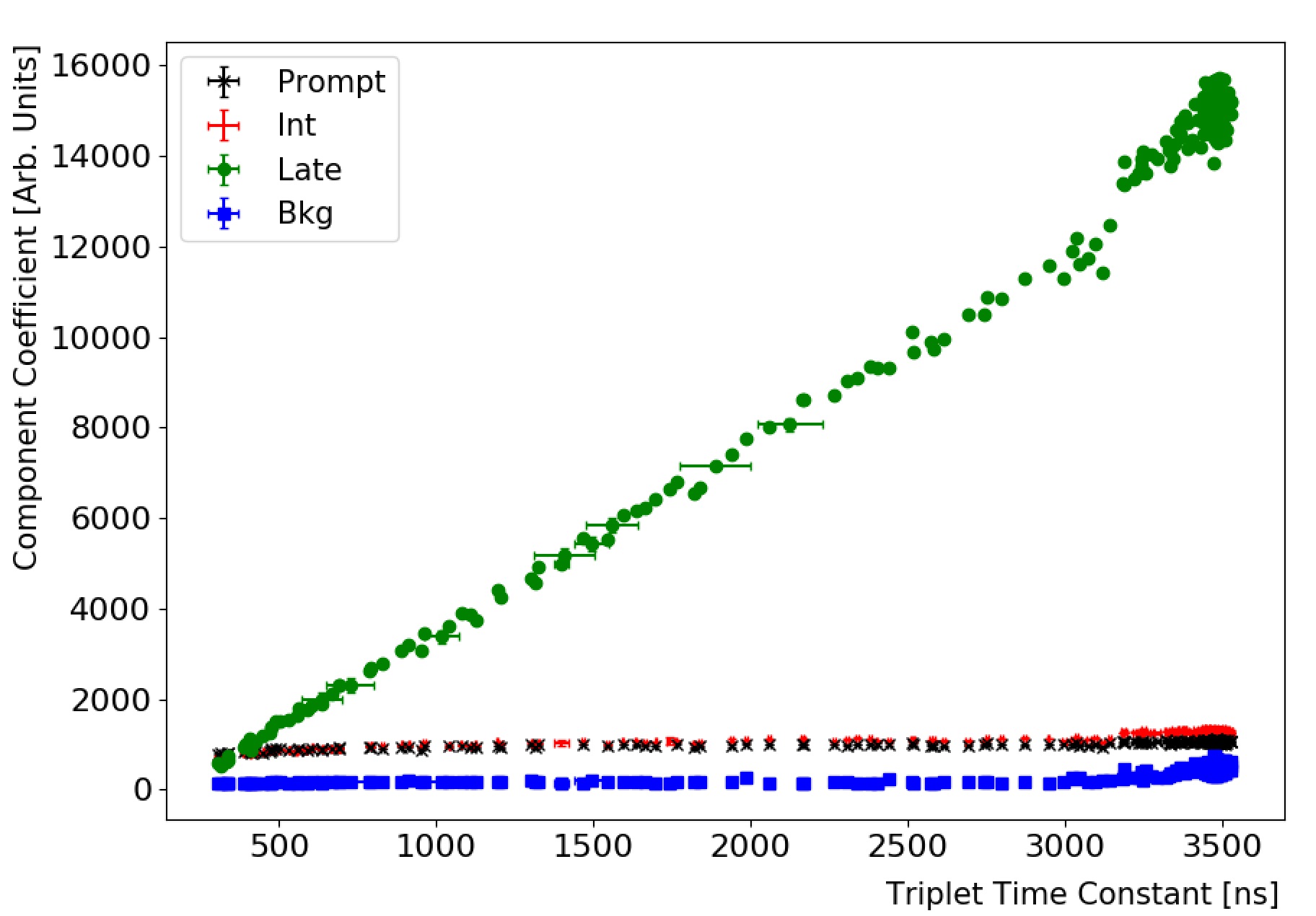}
\caption{\label{fig:fmeanpe}    Mean light yield for each component vs triplet lifetime.  }
\end{figure}

 The PMT photon time-of-flight corrected timing distribution is fitted with the following functional form, consisting of three exponentials convolved with a Gaussian response function:
\begin{equation}\label{eq:3}
f = A + G(t,\mu,\sigma) \otimes [B \cdot e^{-\frac{t}{\tau_{1}}} +C\cdot e^{-\frac{t}{\tau_{2}}}+ D \cdot e^{-\frac{t}{\tau_{3}}}].
\end{equation}
where $G$ is the gaussian response function, the parameters $\tau_{1}$ ,
$\tau_{2}$ and $\tau_{3}$ are the time constant of the fast, intermediate and
slow-decaying (late) states respectively and the parameters $B$ ,$C$ and $D$ are the
fractions of prompt, intermediate and late states respectively.
Finally, $A$ is the fraction of flat background. An example of the fit is shown
in Fig. \ref{fig:ftrip2}. The mean light yield (LY) for each component was determined from
the fitting parameters (B, C, and D) in Eq. \ref{eq:3} and plotted against triplet
lifetime as shown in Fig. \ref{fig:fmeanpe}. As can be seen, the prompt and
intermediate components are relatively flat compared to the late component. It is
confirmed that the singlet state of the second continuum (intermediate component)
is not affected by the impurities, while the triplet state of the second
continuum (late component) is strongly quenched. Moreover the prompt states,
which mostly come from the third continuum, are not affected by impurities
either.\par

\begin{figure}
\includegraphics[scale=0.25]{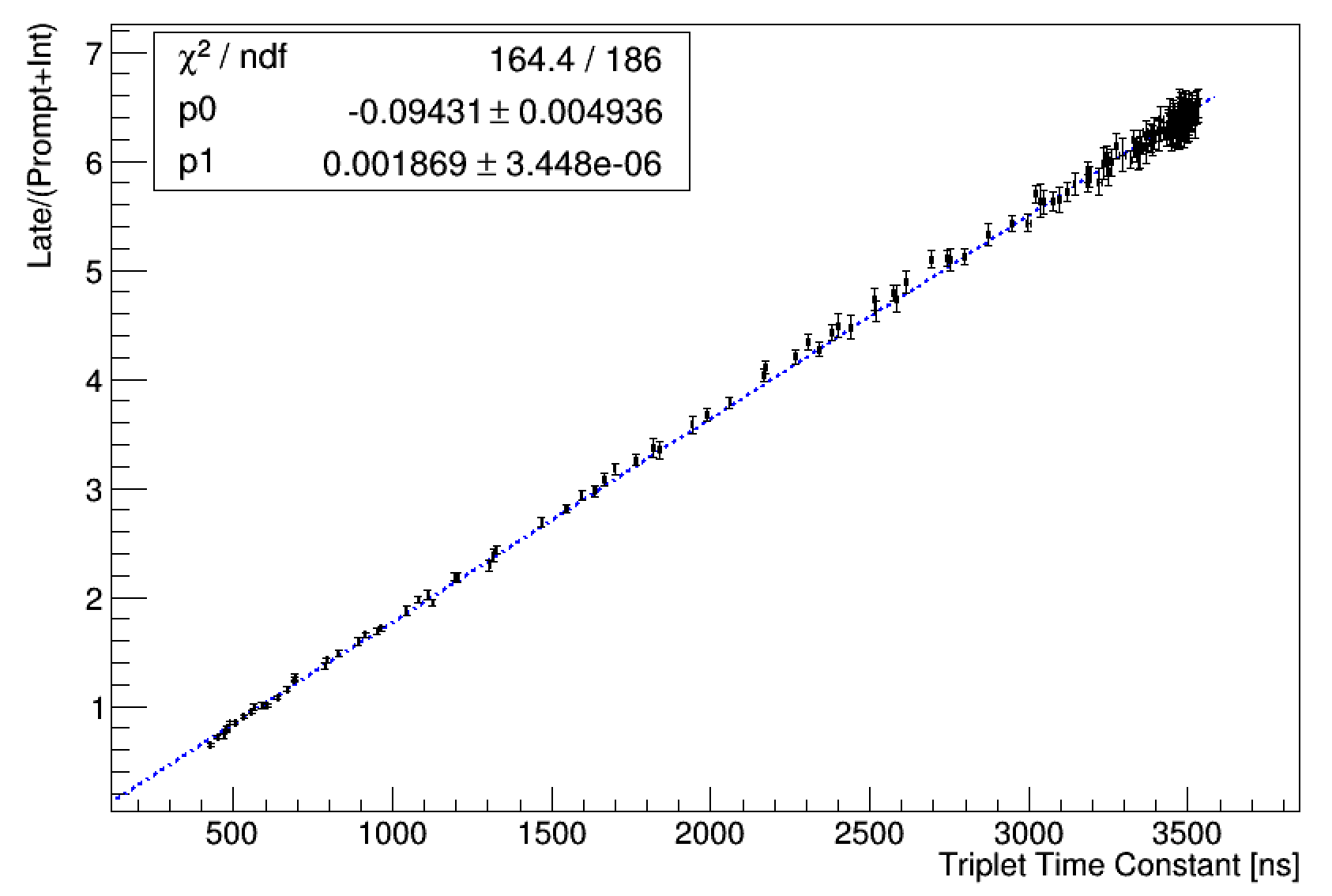}
\caption{\label{fig:flpratiosum} Ratio of late to prompt and intermediate components determine from sum of waveform vs triplet lifetime.}
\end{figure}

The fraction of late component to the total scintillation light is another interesting quantity, which was obtained by taking the fitting parameters in Eq. \ref{eq:3} ($D$/($B$+$C$)). The final results as a function of the average triplet lifetime hour by hour as shown in Fig. \ref{fig:flpratiosum}. The measured ratio is fit to a first order polynomial function. The largest late/prompt ratio, 6.37 $\pm$ 0.01, is determined by averaging the value obtained with long lifetime ($>$ 3.4 $\mu$s). We noticed that  a systematic bump in late component in Fig. \ref{fig:fmeanpe} is correlated with the turning on of additional PMTs which are not included in this analysis. The displacement in the late component is likely due to the increased noise levels (e.g. PMT cross talk), and this effect appears to approximately cancel out when evaluating the ratio. The discrepancy between our results and the value 5.5 ~$\pm$~0.6 reported in reference \cite{1748-0221-3-02-P02001} is largely due to their use of an alpha particle source, well known to produce more prompt light compared to our electronic recoil events.  Also a longer lifetime is attained in our measurement, which effectively increases the late/(prompt+intermediate) ratio. The authors believe the source of the intermediate component to be a combination of PMT effects (double and late pulsing), delayed singlet state light from the second continuum, and possibly delayed TPB light \cite{PhysRevC.91.035503}. No effort is made to disentangle these effects as the relative fraction of the late component is the primary result of this work. However, these effects are considered as a source of systematic uncertainty.


\subsection{Systematic Uncertainties}
\begin{figure}
\includegraphics[scale=0.28]{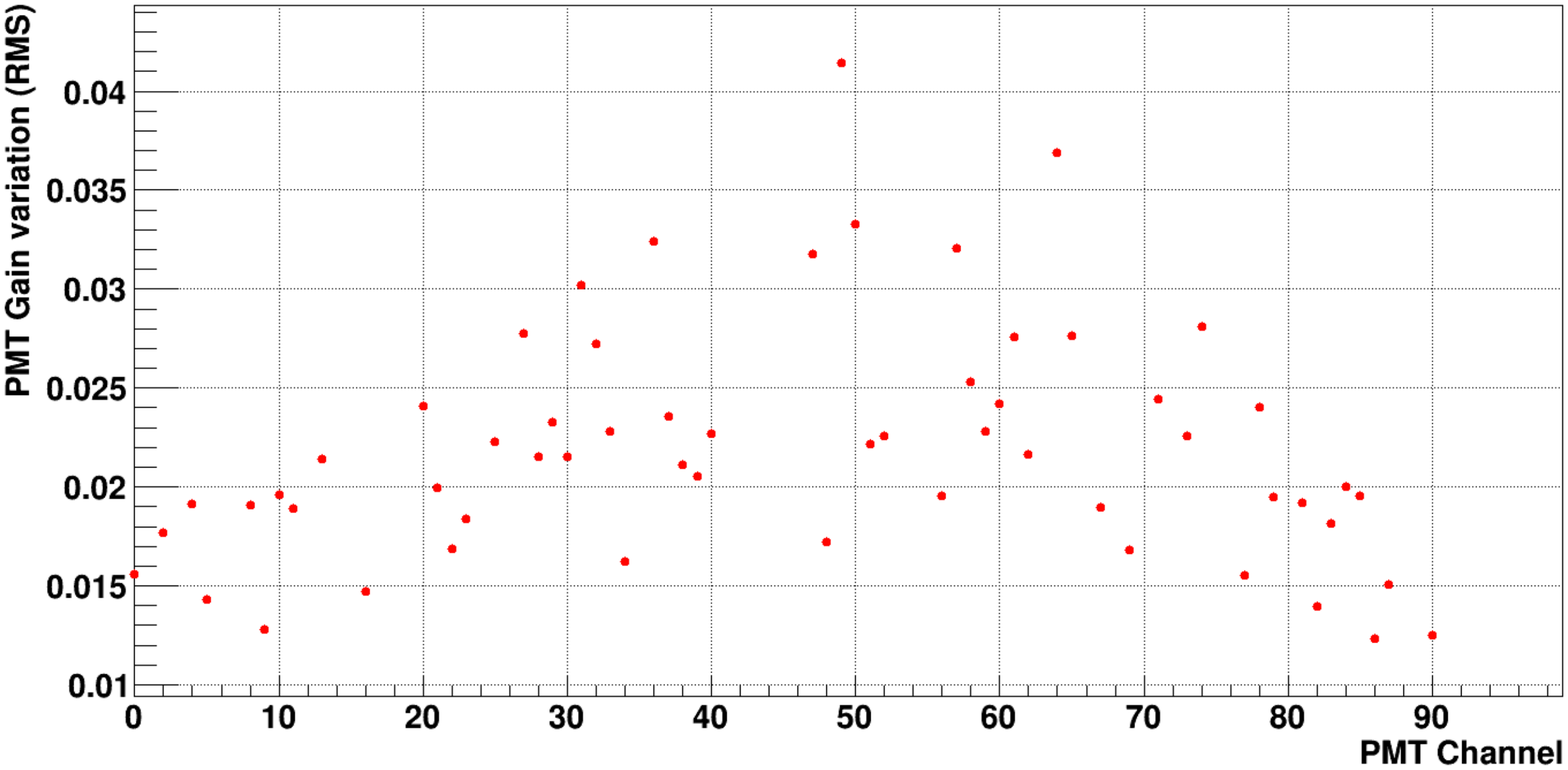}
\caption{\label{fig:gainvar} PMT relative RMS gain variation in Run 962.}
\end{figure}

We have identified a number of systematic effects on the measurement of the triplet lifetime and light yield and estimated uncertainties due to these effects.
The pressure and temperature of IV changed during the measurement,the average pressure and temperature during each cycle was used to determine the density of the gaseous argon. Then by using the equation from P. Moutard's\cite{doi:10.1063/1.452869} (green dashed line in Fig. \ref{fig:ftable}), the systematic uncertainty from the density variation can be determined. The pressure during the pump and purge cycle changed from 1420 mbar to 1560 mbar while the temperature change is much less than 1K, thus the density effect has a negligible systematic uncertainty. The gain variation during the measurement also leads to systematic uncertainties  on triplet lifetime. We determined this uncertainty by tracking the light yield of prompt light hour by hour. Figure \ref{fig:gainvar} shows the relative gain variation during on long run (33 hrs), this run is used to estimate the systematic uncertainty due to the variations of PMT gain. The relative gain variation is defined as root-mean-square (RMS) of the gain variation in 33 hrs divided by mean gain in this 33 hrs. Subsequently, we fit the pulse-time distribution generated by the modified gain (according to the gain variations) and compare with unmodified results. This determines the uncertainty due to gain variations in Table \ref{tab:syslifetime}.\par

\begin{figure}
\includegraphics[scale=0.45]{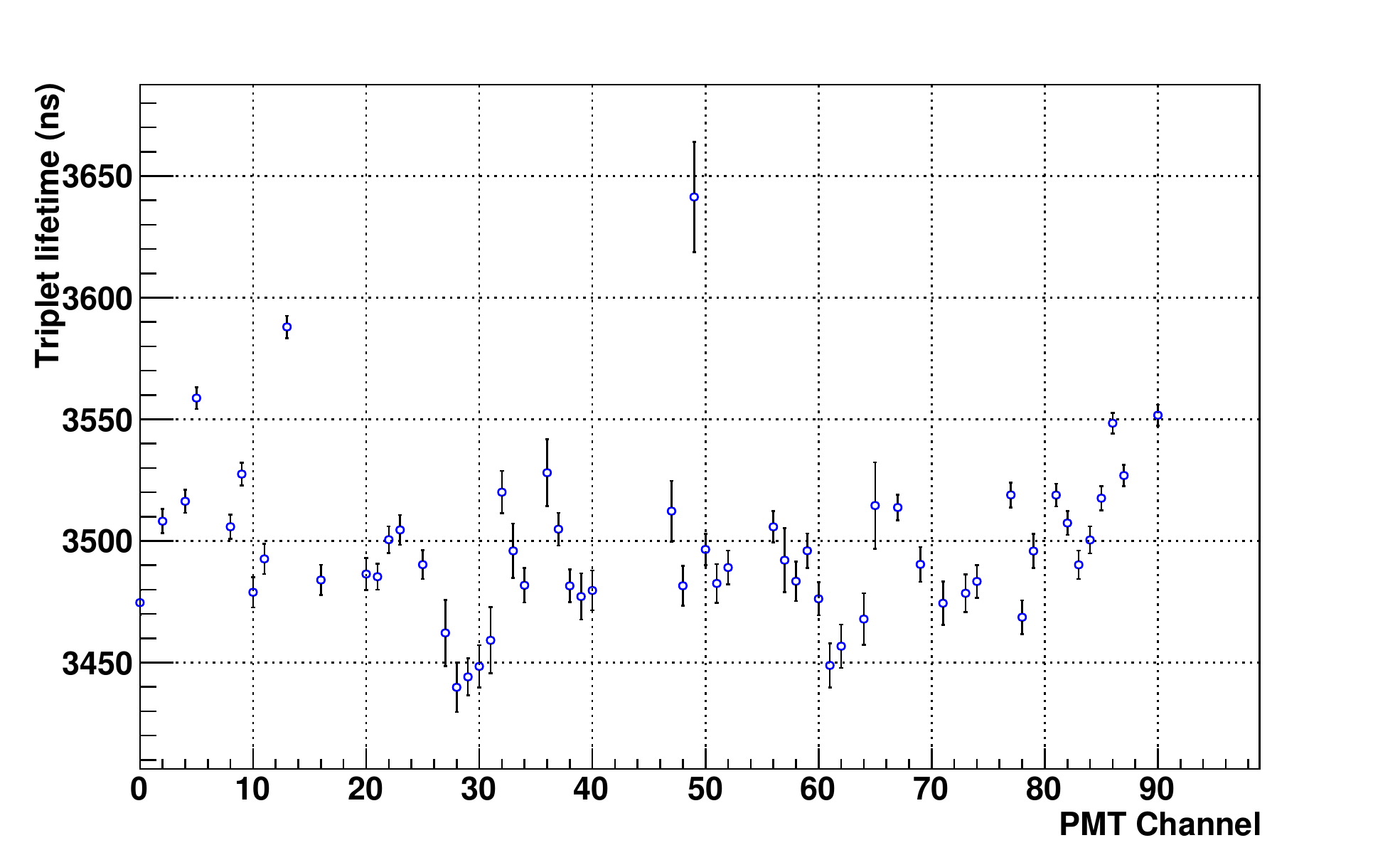}
\caption{\label{fig:tripvar} Triplet lifetime versus PMT.}
\end{figure}

\begin{figure}
\includegraphics[scale=0.45]{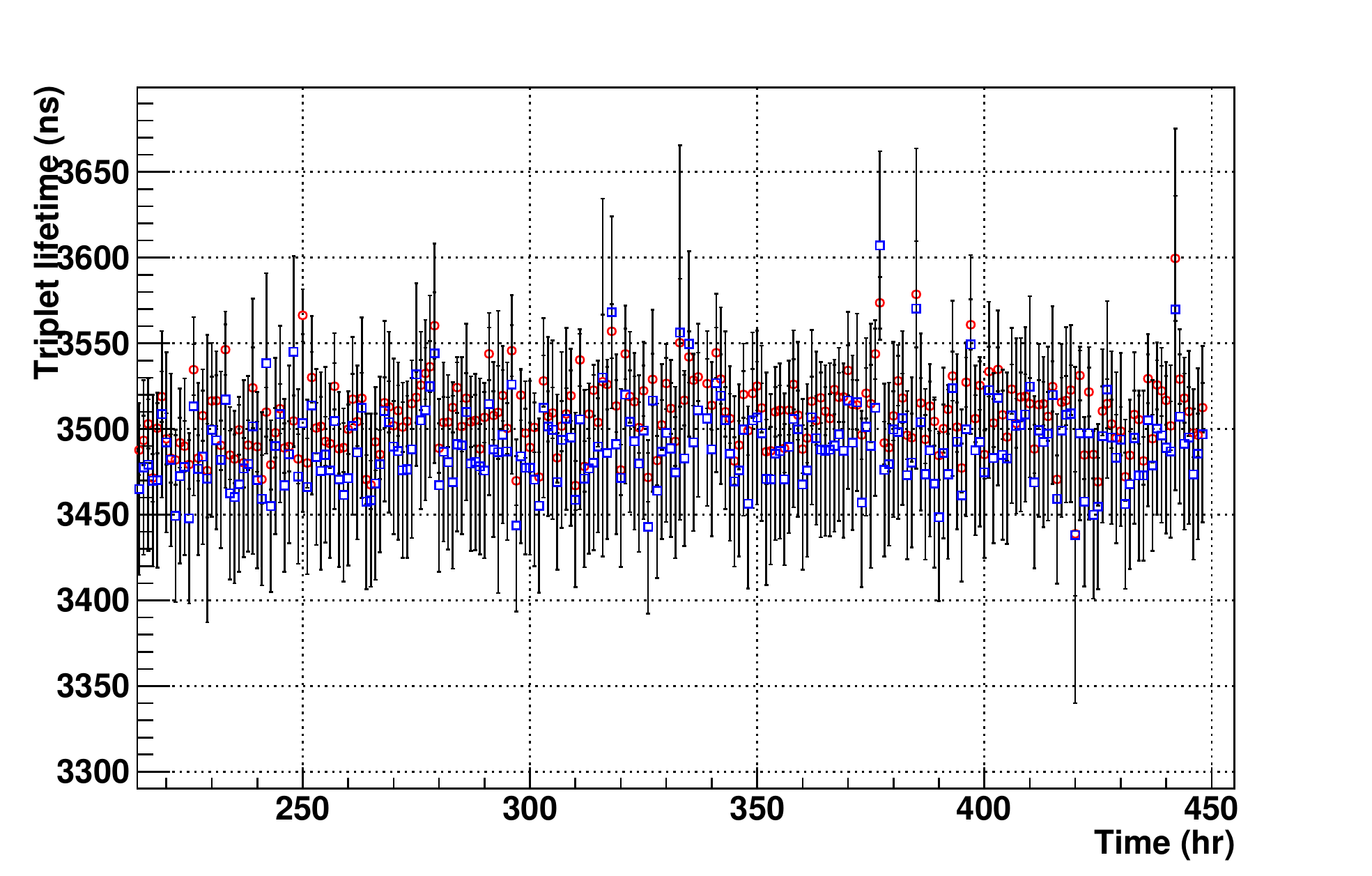}
\caption{\label{fig:tripsumm} Fitting results from pulse-time distribution (red) and sum of the wave- form (blue) overlap in the plot.}
\end{figure}

\begin{figure}
\includegraphics[scale=0.28]{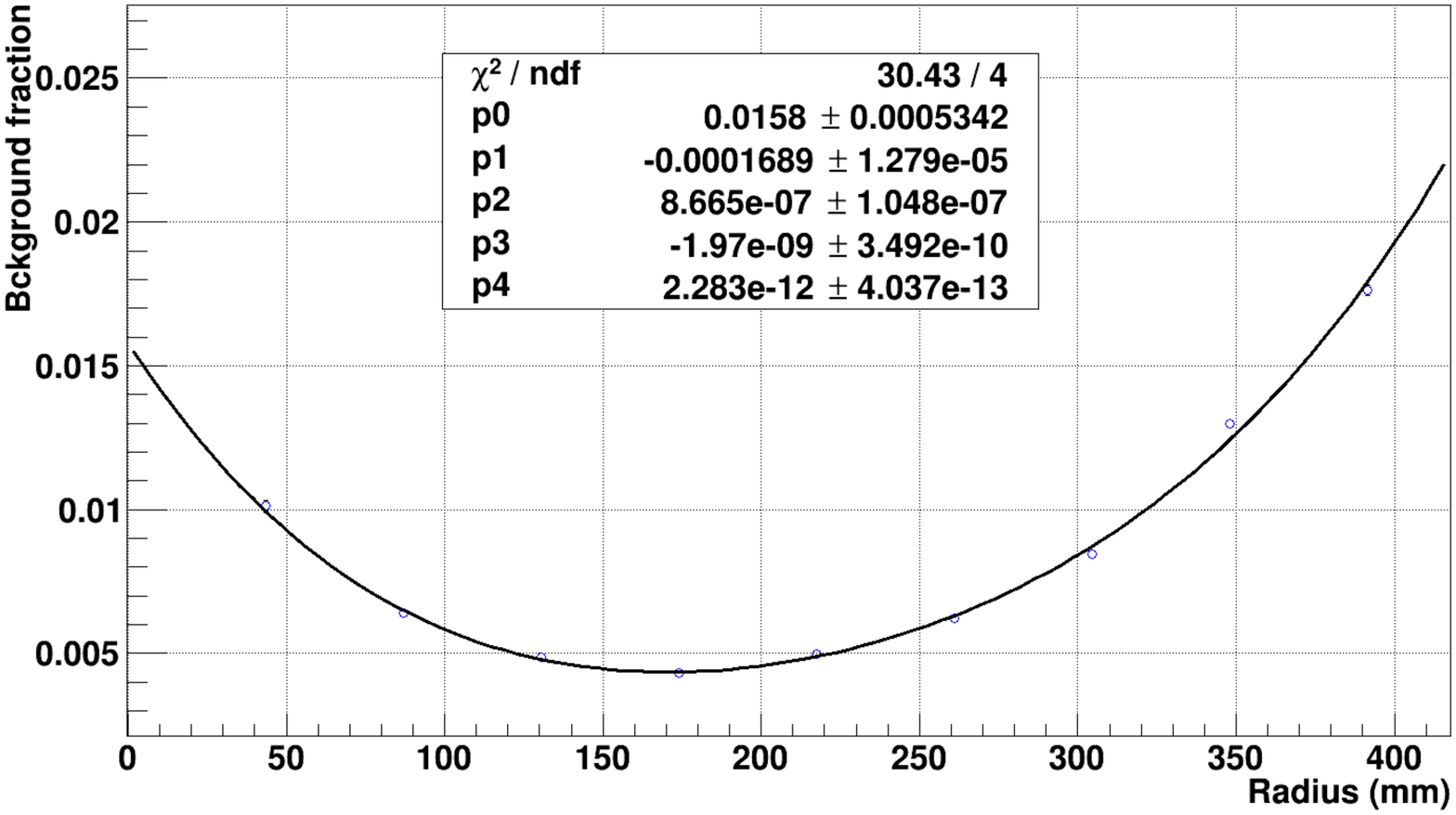}
\caption{\label{fig:rbkg} Fitted background fraction versus radius.}
\end{figure}

On the other hand, due to the potential of inhomogeneity of the gas distribution inside the IV, some PMTs might detect more photons than others. This uncertainty is assessed by fitting the triplet lifetime PMT by PMT. We found that the top PMTs (PMT channel number 1 - 20) has slightly longer triplet lifetime than others. This may be due to the gas non-uniformity at the top of the IV during pumping. The fitted triplet lifetime for each PMTs is shown in Fig. \ref{fig:tripvar}. The systematic uncertainty of pulse finding algorithm can be assesed by comparing the fitting results between pulse-time distribution and summed waveform. The comparison of the fitting results is shown in Fig. \ref{fig:tripsumm}. The estimated systematic uncertainty from pulse finding algorithm is 0.39$\%$. Furthermore, the effect from pumping the detector can be assesed by comparing the triplet lifetime during the pumping and purge (static) respectively. The systematic uncertainty from pumping the detector is estimated to be 0.73 $\%$. Backgrounds also tend to reconstruct near the edge of IV compared to the scintillation events. We developed a pseudo-experiment to investigate this effect. In this method, we fix the triplet time to some constant and varies the background fraction according to the fitted background fraction from different radius (Fig. \ref{fig:rbkg}). Starting with an assumed true triplet lifetime value, we pick a random radius according to Fig. \ref{fig:fnumber}, and obtain the corresponding background fraction from Fig. \ref{fig:rbkg} (interpolating with a 4th order polynomial). Then by sampling the background fraction at different radii, we refit with the fitting function (Eq. (\ref{eq:2})) and use the subsequent variation as the systematic uncertainty. The PMT after-pulsing is also one of the source which contribute to the systematic uncertainty. The after-pulsing of PMTs is simulated by importing the dedicated measurement. Thus, by simulating the detector response with and without PMT after-pulsing, the systematic uncertainty can be accessed. As a result, 0.06$\%$ is contributed by this effect. \par
\begin{figure}
\includegraphics[scale=0.24]{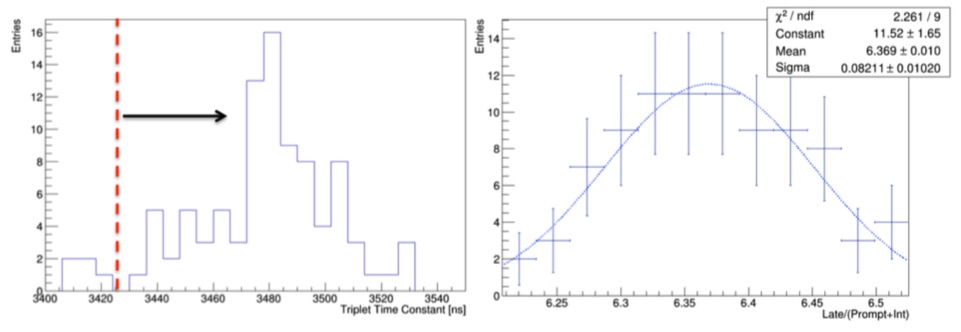}
\caption{\label{fig:ratiouncer1} Profile histograms of the axes in Fig. \ref{fig:flpratiosum} at late triplet time constant. Left: Distribution of measured triplet time constant for large values. Right: Distribution of Late/(Prompt + Intermediate) ratio values after applying a cut of t3 $>$ 3425 ns. The distribution is fit to a Gaussian whose width is taken to be the systematic uncertainty in the spread in the ratio result.}
\end{figure}
\begin{figure}
\includegraphics[scale=0.28]{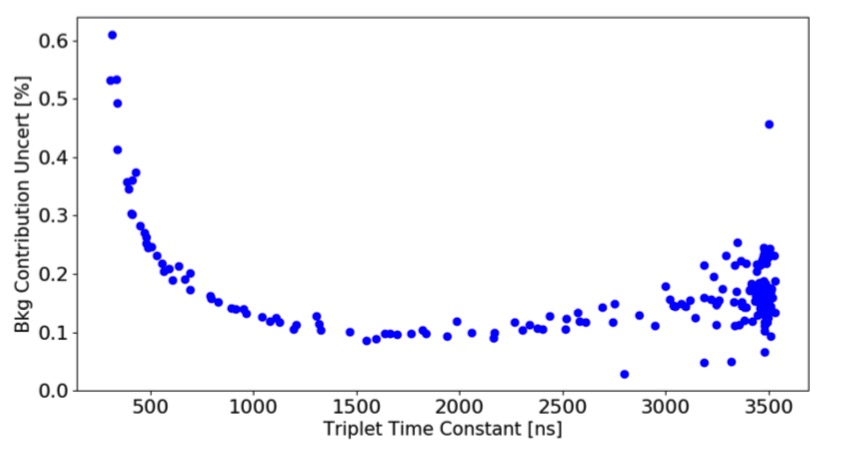}
\caption{\label{fig:ratiouncer2} The percent uncertainty in the Late/(Prompt + Intermediate) ratio as a function of measured triplet time constant when adding and subtracting 1 $\sigma$ of the background coefficient to the late coefficient.}
\end{figure}

Two systematic uncertainties are considered for the measurement of light yield. The first is the spread in the Late/(Prompt + Intermediate) ratio at long triplet time constant values. Evaluating this number provides a handle on the spread in the ratio result once the measured triplet time constant has stabilized. Fig. \ref{fig:ratiouncer1} shows profile histograms of the triplet time constant (left) and \\
Late/(Prompt + Intermediate) (right) ratio axes plotted in Fig. \ref{fig:flpratiosum} for measured triplet time constant between 3400 ns and 3550 ns. Once the triplet time constant measurement stabilizes to 3480 ns, the distribution in the left-side histogram of Fig. \ref{fig:ratiouncer1} motivates placing a cut at 3425 ns to define the population of events from which to evaluate the spread in the ratio. The right-side histogram in Fig. \ref{fig:ratiouncer1} is the distribution of measured Late/(Prompt + Intermediate) ratios after a cut of $>$ 3425 ns on the measured triplet time constant is applied. This distribution is then fit to a Gaussian and the fitted standard deviation is taken to be the systematic spread in the evaluation of the Late/(Prompt + Intermediate). The second systematic uncertainty considered is related to the fitted background contribution. Fig. \ref{fig:ratiouncer2} shows the percent uncertainty as a function of measured triplet time constant in the evaluation of the Late/(Prompt + Intermediate) ratio when adding/subtracting the uncertainty in the fitted background coefficient from the late coefficient. 


\begin{table}
\caption{\label{tab:syslifetime}The source and associated systematic
uncertainties .  }
\begin{tabular}{cccc}
\bf Source & Triplet lifetime & Ratio of LY\\
\hline\hline
PMT gain variation & $\pm$ 0.26\% &--\\
PMT after-pulsing & $\pm$ 0.06\% &--\\
Background & $\pm$ 1.31\% &  $\pm$ 0.20\%\\
PMT variation &  $\pm$ 0.96\% &-- \\
Detector &-- &  $\pm$ 1.25\%\\
Density & $\pm$ 0.07\% & --\\
Pumping on the IV &  $\pm$ 0.73\% &-- \\
Pulse finding & $\pm$ 0.39\%& --\\
\bf Total & \bf $\pm$ 1.84\% & \bf $\pm$ 1.27\%\\
\hline\hline
\end{tabular}
\end{table}



\section{Conclusion}
\label{S:4}
Triplet lifetime is important for performing pulse shape discrimination in
gaseous/liquid argon detectors. Residual impurities in argon will lead to the
degradation of both triplet lifetime and light yield. This subsequently affects
the pulse shape discrimination power and the energy resolution of the detector.
We have investigated the detailed correlation between triplet lifetime and
impurity level, which helps to monitor detector health before, during and after
the commissioning of the detector. We have found the triplet lifetime with
$<$ppb impurity is 3.480 $\pm$ 0.001 (stat.) $\pm$ 0.064 (sys.) $\mu$s, which is the longest lifetime so
far measured at 1.5 bar and temperature less than 140 K, to our knowledge. \par
We have presented the results of relative light yield and the
late/(intermediate + prompt) ratio measurements. The relative light yield
reveals the origin of the quenching effects of impurity molecules. They mainly
interact with the triplet state of the second continuum and results in a loss
of light yield. A detailed analysis of light yield from each component suggest
that the impurity molecules did not affect the prompt (third continuum) and
singlet state (second continuum) except at very high impurity levels ($>$100
ppm). This is probably due to the relatively fast lifetime of these two
components, and the low reaction rates when impurity levels are low. The
late/prompt ratio gives a quantitative definition of the scintillation timing
structure. The best value of the late/(intermediate + prompt) ratio has been
measured as 6.37 $\pm$ 0.01(stat.) $\pm$ 0.08 (sys.), which has better precision than previously
measured\cite{1748-0221-3-02-P02001} due to the high purity of our argon, as
evidenced by our long triplet lifetime. 


\section*{Acknowledgments}
This research is supported by Pacific Northwest National Laboratory, SNOLAB and DOE (Office of Science, DE-FG02-04ER41300). We gratefully acknowledge the support of the U.S. Department of Energy through the \\
LANL/LDRD Program for this work. This work also supported by University of California, Berkeley and the Hellman Faculty Fellowship Fund.

\bibliographystyle{unsrt}

\end{document}